\documentclass[usenatbib]{mnras}
\usepackage{natbibmnfix, graphicx, times, amsmath, epsfig, amssymb}
\usepackage{verbatim}
\usepackage{color}
\usepackage{mathptmx}
\usepackage{txfonts}
\usepackage[T1]{fontenc}
\usepackage{aecompl}

\def\simgt{\lower.5ex\hbox{\gtsima}} 
\def\simlt{\lower.5ex\hbox{\ltsima}} 
\def\gtsima{$\; \buildrel > \over \sim \;$} 
\def\ltsima{$\; \buildrel < \over \sim \;$}
\def\Msun{M_\odot}

\newcommand\lsim{\mathrel{\rlap{\lower4pt\hbox{\hskip1pt$\sim$}}
        \raise1pt\hbox{$<$}}}
\newcommand\gsim{\mathrel{\rlap{\lower4pt\hbox{\hskip1pt$\sim$}}
        \raise1pt\hbox{$>$}}}
\def\myputfigure#1#2#3#4#5%
{\vskip#5pt\makebox[0pt]{\hskip#2in
\includegraphics[width=#3\textwidth]{#1}}\vskip#4pt\hfill}

\title[The Growth Efficiency of High-Redshift Black Holes]
      {The Growth Efficiency of High-Redshift Black Holes}
\author[F. Pacucci et al.]
{Fabio Pacucci$^1$ \thanks{fabio.pacucci@sns.it},
Marta Volonteri$^{2}$, Andrea Ferrara$^{1,3}$\\
$^1$Scuola Normale Superiore, Piazza dei Cavalieri, 7  56126 Pisa, Italy \\
$^2$Institut d'Astrophysique de Paris, Sorbonne Universit\'{e}s, UPMC Univ Paris 6 et CNRS, UMR 7095, 98 bis bd Arago, 75014 Paris, France \\
$^3$Kavli Institute for the Physics and Mathematics of the Universe (WPI), Todai Institutes for Advanced Study, \\ 
\, the University of Tokyo 5-1-5 Kashiwanoha, Kashiwa, 277-8583, Japan \\
}
             
\date{Accepted for publication in MNRAS}

\begin{document}
\label{firstpage}
\pagerange{\pageref{firstpage}--\pageref{lastpage}}
\maketitle
             
\begin{abstract}
The observational evidence that Super-Massive Black Holes ($M_{\bullet} \sim 10^{9-10} \, \mathrm{\Msun}$) are already in place less than $1 \, \mathrm{Gyr}$ after the Big Bang poses  stringent time constraints on the growth efficiency of their seeds.
Among proposed possibilities, the formation of massive ($\sim 10^{3-6} \, \mathrm{\Msun}$) seeds and/or the occurrence of super-Eddington ($\dot{M}>\dot{M}_{Edd}$) accretion episodes may contribute to the solution of this problem.
In this work, using a set of astrophysically-motivated initial conditions, we analytically and numerically investigate the accretion flow onto high-redshift ($z \sim 10$) black holes to understand the physical requirements favoring rapid and efficient growth.
Our model identifies a ``feeding-dominated" accretion regime and a ``feedback-limited" one, the latter being characterized by intermittent (duty cycles ${\cal D} \lsim 0.5$) and inefficient growth, with recurring outflow episodes. 
We find that low-mass seeds ($\lsim 10^{3-4} \, \mathrm{\Msun}$) evolve in the feedback-limited regime, while more massive seeds ($\gsim 10^{5-6} \, \mathrm{\Msun}$) grow very rapidly as they are found in the feeding-dominated regime.
In addition to the standard accretion model with a fixed matter-energy conversion factor ($\epsilon = 0.1$), we have also explored slim disk models, appropriate for super-Eddington accretion, where radiation is trapped in the disk and the radiative efficiency is reduced ($\epsilon \lsim 0.04$), which may ensure a continuous growth with $\dot{M} \gg \dot{M}_{Edd}$ (up to $\sim 300\dot{M}_{Edd}$ in our simulations). Under these conditions, outflows play a negligible role and a black hole can accrete $80\%-100\%$ of the gas mass of the host halo ($\sim 10^7 \, \mathrm{\Msun}$) in $\sim 10 \, \mathrm{Myr}$, while in feedback-limited systems we predict that black holes can accrete only up to $\sim 15\%$ of the available mass.
\end{abstract}


\setcounter{footnote}{1}
\newcounter{dummy}
\section{Introduction}
\label{sec:introduction}
Recent observations \citep{Mortlock_2011, Wu_2015} have detected the presence of optically bright quasars at redshifts as high as $z \sim 7$. These high-energy sources are powered by accretion onto Super-Massive Black Holes (SMBHs), suggesting the presence of compact objects with mass $M_{\bullet} \sim 10^{9-10} \, \mathrm{\Msun}$ less than $1 \, \mathrm{Gyr}$ after the Big Bang \citep{Fan_2006}.
This evidence contrasts with the standard theory of black hole growth, which requires a longer time to build up such massive objects (see \citealt{Haiman_2013} for a recent review), due to: (i) the low mass of some of the proposed seeds, born out of first-generation (Pop III) stars, with masses $M_{\bullet} \lsim 10^3 \, \mathrm{\Msun}$ \citep{Madau_Rees_2001, Bromm_Loeb_2003, Petri_2012}, and (ii) the maximum growth rate allowed for a radiatively efficient and spherical inflow, the Eddington rate, which provides a lower limit for the time scale of the process \citep{Jeon_2012}. 
Generally speaking, the luminosity $L$ emitted due to a gas inflow with an accretion rate $\dot{M} \equiv dM/dt$ is $L=\epsilon c^2 \dot{M}$, where $\epsilon$ is the matter-radiation conversion factor and $c$ is the speed of light.
In the standard scenario, a black hole grows in mass exponentially, with a time scale given by the Salpeter time $t_{S} \sim 0.045\epsilon_{0.1} \, \mathrm{Gyr}$, where $\epsilon_{0.1}$ is normalized to the standard value of $10\%$: starting from a low-mass seed ($M_{\bullet} \sim 100 \, \mathrm{\Msun}$), this process would require a \textit{constant} accretion at the Eddington rate up to $z \sim 7$ to build up a $\sim 10^9 \, \mathrm{\Msun}$ SMBH.

Several ways to overcome these limitations have been proposed in the literature.
The possibility of giving a jump start to the growth process through more massive seeds has been investigated thoroughly (see e.g. \citealt{Volonteri_2010}) with a variety of mechanisms: (i) the direct collapse of self-gravitating pre-galactic disks at high-redshifts (\citealt{Lodato_Natarajan_2006,Begelman_2006, Lodato_2007}), (ii) the formation of a very massive star from runaway stellar mergers in a dense cluster (\citealt{Devecchi_2009, Davies_2011}) and (iii) the collapse of a primordial atomic-cooling halo, exposed to a Lyman-Werner flux of sufficient intensity, into a Direct Collapse Black Hole (DCBH), through a general relativistic instability (\citealt{Shang_2010, Johnson_2012, Ferrara_2014}).

An alternative scenario assumes that accretion rates are not capped by the Eddington limit \citep{Volonteri_2005}.
Recent works (\citealt{Alexander_2014, Madau_2014, Volonteri_2014}) have proposed the occurrence of short and recurring, but strongly super-critical (i.e. super-Eddington) accretion episodes at high-redshifts, with rates as large as $50-100$ times the Eddington limit $\dot{M}_{Edd} \equiv L_{Edd}/(\epsilon c^2)$, where:
\begin{equation}
 L_{Edd} \equiv \frac{4 \pi G M_{\bullet} c}{\kappa_{T}} \, .
 \label{L_Edd_def}
\end{equation}
In the definition of the Eddington luminosity, $G$ is the gravitational constant and $\kappa_{T}$ is the Thomson opacity.
The radiative efficiency of the gas inflow depends on the accretion rate. 
Recalling the definition of the Eddington luminosity (Eq. \ref{L_Edd_def}), we define $f_{Edd} \equiv \dot{M_{\bullet}}/\dot{M}_{Edd}$, i.e. the accretion rate normalized to the Eddington value.
It is expected that if matter is accreted at moderate rates $0.01 \lsim f_{Edd} \lsim 1$, the inflowing material creates a radiatively efficient, geometrically thin and optically thick accretion disk \citep{Shakura_Sunyaev_1973}.  In this case the radiative efficiency depends only on the black hole spin, and varies from $\sim 6\%$ for Schwarzschild black holes to $\sim 32\%$ for maximally rotating ones \citep{Thorne_1974}. 
If, instead, accretion occurs super-critically ($f_{Edd} >1$) the structure of the accretion disk is modified because of advection: the energy produced in the disk is carried inwards, in the black hole, rather than being radiated away \citep[see, e.g.,][]{2013LRR....16....1A,2015arXiv150502172L}. The disk thickness increases and the disk becomes geometrically thick. The most common solution proposed for such accretion flows is the ``slim disk" (\citealt{Abramowicz_1988,Paczynski_1982,Mineshige_2000,Sadowski_2009,Sadowski_2011,McKinney_2014}), radiatively inefficient and with a thick geometric structure, in which photon trapping is significant inside the trapping radius $R_{pt}$:
\begin{equation}
R_{pt} = \frac{R_s}{\epsilon} \frac{\dot{M}}{\dot{M}_{Edd}} \, ,
\end{equation}
where $R_s$ is the Schwarzschild radius.
Only a fraction of the photons produced by the viscous process inside the accretion flow is able to free stream out of the trapping radius, because the photon diffusion time exceeds the time scale for accretion. Consequently, the effective radiation pressure acting on the surrounding gas is decreased (see e.g. \citealt{Begelman_1978, Ohsuga_2002}) and the luminosity is only mildly (e.g., logarithmically) dependent on the accretion rate.
Alternatives of the slim disk solution exist, e.g the ZEro-BeRnoulli Accretion \citep[ZEBRA][]{Coughlin_2014} and the ADiabatic Inflow-Outflow Solutions \citep[ADIOS][]{1999MNRAS.303L...1B,2012MNRAS.420.2912B} models , which allow for a fraction of the inflowing mass to be lost during the accretion process.

The present work investigates, both analytically and numerically, the growth of high-redshift ($z \sim 10$) black hole seeds.
The growth process may be controlled by the amount of gas available in the halo (feeding-dominated) or by the radiative feedback (feedback-limited).
The growth is feeding-dominated if the radiative back-reaction of the black hole is negligible: the rapidity of the process is mainly determined by the gas accretion rate that the host halo can provide.
The black hole growth is more efficient and rapid if the flow is feeding-dominated, assuming that a sufficient amount of gas is present in its host halo.
  
We devise a very general analytic model which is able to predict the growth efficiency from the physical properties of the system formed by the black hole seed and its host halo.
Furthermore, we employ a radiation-hydrodynamic code to follow the growth process  from small ($0.002$ pc) to large scales ($\gg R_B$, where $R_B$ is the Bondi radius), spanning a spatial dynamic range of four orders of magnitude, with special emphasis on the properties of the inner regions of the host halo, providing most of the accretion material to the central object.
The growth is monitored as a function of different parameters, namely: the accretion model (radiatively efficient or inefficient), the density profile of the halo $\rho(r)$ and the initial mass of the seed $M_0=10^{3-6} \, \mathrm{\Msun}$.

This study is a natural follow-up of the paper by \cite{Pacucci_2015}, where the authors have simulated in great detail the accretion process onto a $z \sim 10$ black hole seed of initial mass $10^5 \, \mathrm{\Msun}$, embedded in a dark matter halo with a gas mass of $\sim 10^7 \, \mathrm{\Msun}$ and with extreme density conditions (a number density of hydrogen particles at $\sim 0.1 \, \mathrm{pc}$ from the center of the halo $\sim 10^7 \, \mathrm{cm^{-3}}$), finding that in $\sim 142 \, \mathrm{Myr}$ about $90\%$ of the gas mass of the halo has been accreted onto the compact object.

In this general framework, we aim at clarifying several aspects, including (i) do the radiatively inefficient accretion models provide an effective way to rapidly increase the mass of the seed? If so, why? (ii) What is the final fate of black hole seeds as a function of their initial mass? How long can they accrete? (iii) Is it possible to predict the growth efficiency of the accretion process from the physical properties of the (black hole + host halo) system?
In the cosmological context, answering these questions will provide some insights into the mass that has been locked inside black holes during the cosmic evolution. This would be of great interest to understand both the formation of high-redshift SMBHs and their remnant population.

The outline of this paper is as follows. In $\S 2$ we briefly describe the physics and equations of the radiation-hydrodynamic problem, along with the initial conditions for the density profiles. In $\S 3$ we present our analytic model for the black hole growth, while in $\S 4$ we show the results of our simulations. Finally, in $\S 5$ we provide some further discussion and the conclusions.
Throughout, we adopt recent Planck cosmological parameters \citep{Planck_2015}: $(\Omega_m, \Omega_{\Lambda}, \Omega_b, h, n_s, \sigma_8 )= (0.32, 0.68, 0.05, 0.67, 0.96, 0.83)$.

\section{Physical and numerical implementation}
\label{sec:methods}
The present study employs a series of radiation-hydrodynamic simulations to test the predictions of our growth model. Our code is designed to perform a fully consistent treatment of uni-dimensional spherically-symmetric hydrodynamic equations and a simplified, frequency-integrated, version of radiative transfer equations. 
While the spherical symmetry is an idealization of a real accretion flow, several works have shown that 1D simulations provide a reliable description of many of its most important features.
For instance, in \cite{Novak_2011} the authors performed a comparative analysis between the outputs of a code run in 1D and 2D, finding similar results in terms of black hole growth and duty cycle. The main difference, for which the multi-dimensional approach is a significant improvement, concerns the fact that the additional degrees of freedom allow classical instabilities (e.g. the Rayleigh-Taylor and the Kelvin-Helmholtz ones) to operate. Their net effect is to produce a somewhat higher accretion rate, a less effective feedback and a more irregular pattern of bursts, compared with the 1D case.
Notwithstanding the dimensionality of the code, our implementation cannot take into account the full complexity of the accretion flow, which only a more advanced treatment of angular momentum transport would allow.
Due to the triaxiality of the host halo, the angular momentum field is extremely complex and variable in time, at every location inside the inflow (see \citealt{Choi_2013, Choi_2015}). Overall, the outward angular momentum transfer is very efficient \citep{Choi_2015}, due to the gravitational torques induced by both the dark matter and the gas distributions. The gas loses its angular momentum efficiently and flows well beyond its centrifugal barrier: the gas that reaches the black hole is expected to have low angular momentum. Therefore, despite its simplifications, our approach helps in acquiring physical insight on the process of black hole growth. In this Section we present a general overview of the most important aspects of the physical implementation, while the interested reader is referred to the paper \cite{Pacucci_2015} for a detailed description of the code.

The domain of our simulations\footnote{We performed a series of convergence tests on the extension of the spatial range (see \citealt{Pacucci_2015}) which confirmed that the main outputs of our simulations (e.g. duty cycles and accretion rates) do not depend on it as long as (i) it covers a sufficiently large radial range around $R_B$ (e.g. from $\sim 0.1 R_B$ to $\sim 2 R_B$) and (ii) the centrifugal radius, where the accretion disk starts to form, is not resolved.} spans approximately from $0.002 \, \mathrm{pc}$ to $20 \, \mathrm{pc}$.
The characteristic spatial scale for accretion is the Bondi radius $R_B$:
\begin{equation}
R_B = \frac{GM_{\bullet}}{c_{s(\infty)}^2} \, .
\end{equation}
Here, $c_{s(\infty)}$ is the sound speed at large distances from the black hole, defined as:
\begin{equation}
c_{s(\infty)} = \sqrt{\frac{\gamma R T_{\infty}}{\mu}} \, ,
\end{equation}
where $\gamma = 5/3$ is the ratio of specific heats, $R$ is the gas constant, $T_{\infty}$ is the gas temperature at large distances and $\mu=1.15$ is the mean molecular weight for a primordial gas. For instance, the Bondi radius corresponding to a seed with initial mass $M_{\bullet} = 10^5 \, \mathrm{\Msun}$ with $c_{s(\infty)} \sim 12 \, \mathrm{km \, s^{-1}}$ (i.e. $T_{\infty}\sim 10^4 \, \mathrm{K}$) is $R_B = 3 \, \mathrm{pc} \sim 10^{-3} \, R_{\rm vir} \sim 10^{8} \, R_{\rm S}$, where $R_{vir}$ is the virial radius of the halo.
Our spatial range, spanning $\sim 4$ orders of magnitude, covers the entire range of $R_B$ corresponding to an initial mass of the seed within the range $10^{3-6} \, \mathrm{\Msun}$.

In the hydrodynamic module we solve the standard system of ideal, non-relativistic Euler's equations (conservation laws for mass, momentum and energy, neglecting viscosity, thermal conduction and magnetic fields) for a primordial (H-He) composition gas with helium fraction $Y_P=0.24665$ \citep{Planck_2015} and no metals, spherically accreting onto a central black hole, supposed at rest and already formed at the beginning of the runs, with a given initial mass $M_0 \equiv M_{\bullet}(t=0)$. The infalling gas has zero angular momentum with respect to the black hole.

The forces acting on the gas are: (i) the thermal pressure, (ii) the gravitational pull of the black hole and (iii) the radiation pressure.
The thermal pressure is given by the usual equation for ideal gases:
\begin{equation}
P_g = \frac{ \rho R T}{\mu} \, ,
\label{EOS}
\end{equation}
where $\rho$ is the mass density.
The gravitational acceleration at the distance $r$ from the central object is:
\begin{equation}
g(r,t) = -\frac{GM_{\bullet}(t)}{r^2} \, .
\end{equation}
The value of the black hole mass $M_{\bullet}(t)$ changes with time, due to gas accretion, with the following set of rules, where $\dot{M}_{\bullet}=4 \pi r^2 \rho |v|$:
\begin{equation}
  \begin{cases}
   \dot{M}_{\bullet}(t) \neq 0 \, \, \,  \Leftrightarrow \, \,\, v(r=r_{0},t)<0 \\
   M_{\bullet}(t) = M_0 + (1-\epsilon)\int_0^t  \! \dot{M}_{\bullet} \, \mathrm{d}t
  \end{cases}
\end{equation}
where $v$ is the velocity of the gas and $\epsilon$ ranges\footnote{We assume that the matter-radiation conversion factor $\epsilon$ is numerically equal to the accretion efficiency $\eta$, i.e. all the gas reaching the inner boundary is actually accreted by the black hole (see e.g. \citealt{Sadowski_2014_b, McKinney_2014, Jiang_2014}).} from $\epsilon = 0.057$ for a Schwarzschild (i.e., non-rotating) black hole to $\epsilon=0.32$ for a maximally rotating object (see \citealt{Thorne_1974}).
The mass flow $\dot{M}_{\bullet}$ is computed at the accretion radius $r_{0}$ (the innermost cell of our spatial grid) and it is equivalent, by definition, to the accretion rate onto the black hole.
The acceleration caused by the radiation pressure is:
\begin{equation}
a_{rad}(r) = \frac{\kappa(\rho, T) L(r)}{4 \pi r^2 c} \, ,
\label{a_rad}
\end{equation}
where $L$ is the emitted luminosity and $\kappa(\rho, T)$ is the opacity of the gas, which includes the Thomson term (with the additional inclusion of a temperature dependence, as in \citealt{Begelman_2008}) and bound-free terms.
In our code, all radiation-related quantities are integrated over frequencies: matter and radiation are coupled via Thomson (electron) scattering and bound-free interactions.
The radiative transfer employs a simplified two-streams approximation presented in \cite{Novak_2012}, to which the interested reader is deferred for a detailed description.

The relation between the emitted luminosity $L_0 \equiv L(r_0)$ and the accretion rate $\dot{M}_{\bullet}$ is:
\begin{equation}
\begin{cases}
L_0 \equiv  \epsilon c^2 {\cal F}(\dot{M_{\bullet}}) & \text{if } v(r_{0}) <0 \\
L_0 \equiv 0 & \text{if } v(r_{0}) \geq 0
\end{cases}
\label{L_definition}
\end{equation}
where ${\cal F}(\dot{M_{\bullet}})$ is a generic function of the accretion rate.
In the simple case of radiatively efficient accretion (treated in \citealt{Pacucci_2015}) the following relation holds:
\begin{equation}
\frac{L}{L_{Edd}} = f_{Edd} \, .
\label{radiatively_efficient}
\end{equation}
In the radiatively inefficient accretion mode (for instance, the slim disk solution) the luminosity $L$ depends on $f_{Edd}$ as in the following prescriptions \citep{Mineshige_2000,Volonteri_2014}:
\begin{equation}
\frac{L}{L_{Edd}} = \frac{f_{Edd}}{25} \, \, \, (f_{Edd}<50) ;
\label{radiatively_inefficient_1}
\end{equation}
\begin{equation}
\frac{L}{L_{Edd}} = 2\left[1+\ln{\frac{f_{Edd}}{50}}\right] \, \, \, (f_{Edd}\geq50) .
\label{radiatively_inefficient_2}
\end{equation}
In this scenario, only a fraction of the emitted luminosity escapes to infinity, and the effective radiation pressure depends only weakly on the accretion rate: note that $L=2L_{Edd}$ for $f_{Edd}=50$. In the radiatively inefficient accretion mode the matter-radiation conversion efficiencies are $\epsilon = 0.04$ for $f_{Edd}<50$ and:
\begin{equation}
\epsilon = \frac{1}{25} \left(\frac{f_{Edd}}{50}\right)^{-1} \left[ 1 + \ln\left(\frac{f_{Edd}}{50}\right) \right] \leq 0.04 \, .
\label{epsilon_SD}
\end{equation}
for $f_{Edd} \geq 50$.

Due to radiation pressure, the accretion may occur in an intermittent manner. Therefore, it is useful to introduce the duty cycle, defined as the fraction of time spent accreting within a given time frame of duration $t_{tot}$:
\begin{equation}
{\cal D} \equiv 1 - \frac{t_{idle}}{t_{tot}} \, .
\label{DC_def}
\end{equation}
Here, $t_{idle}$ is the idle time, i.e. the time spent without accretion taking place.

\subsection{The initial density profile}
In this work we simulate the spherically-symmetric gas accretion onto a black hole seed at the center of a dark matter halo of virial temperature $T_{vir} \sim 10^4 \, \mathrm{K}$ and total mass (dark matter and baryons) $M_h  = 6.7 \times 10^8 \, \mathrm{M}_{\odot}$ at $z=10$. For $r \ll R_{vir}$, as in our computational domain, most of the mass is baryonic. We assume that the gas initially follows the isothermal ($T \sim 10^4 \, \mathrm{K}$) density profile derived from the simulations in \cite{Latif_2013}, well approximated by the functional form: 
\begin{equation}
\rho(r) = \frac{\rho_0}{1+(r/a)^2} \, ,
\label{density_profile}
\end{equation}
where $a$ is the core radius.

\begin{figure}
\vspace{-1\baselineskip}
\hspace{-0.5cm}
\begin{center}
\includegraphics[angle=0,width=0.50\textwidth]{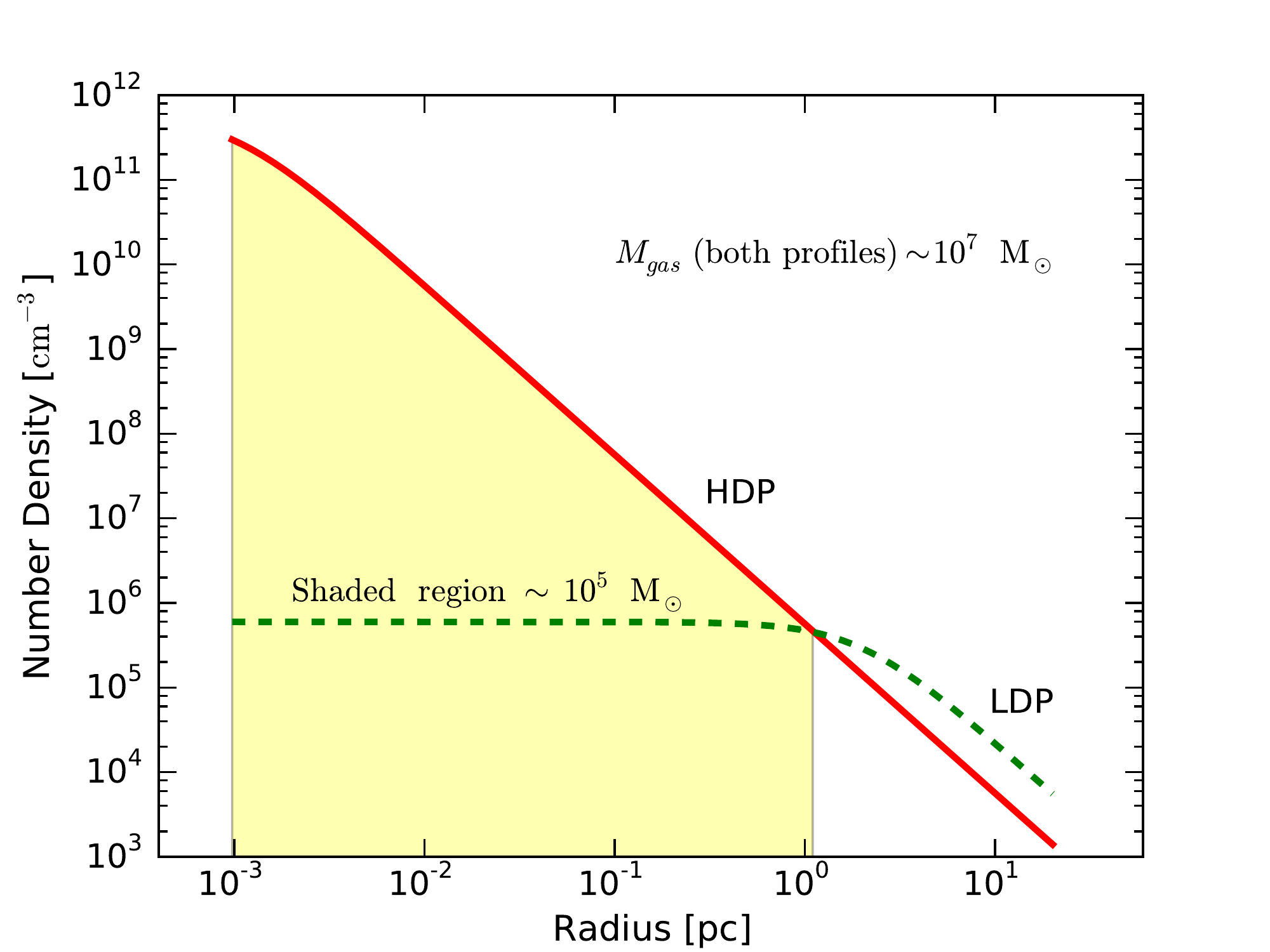}
\caption{The two density profiles employed, both yielding a total baryonic mass of $\sim 10^7 \, \mathrm{\Msun}$ over the integration range. The LDP has a much larger core radius, but the central density is smaller by $\sim 6$ orders of magnitude. The yellow-shaded region contains a total mass of $10^5 \, \mathrm{\Msun}$ and may be interpreted as the gas which has been extracted from the HDP to produce a black hole of the same mass. Consequently, the HDP profile may be interpreted as the density profile resulting after the formation of a small black hole of mass $M_{\bullet} \lsim 10^3 \, \mathrm{\Msun}$.}
\label{fig:density_profiles}
\end{center}
\end{figure}

In order to study how the black hole growth depends on the host halo, we implemented two different density profiles, schematically shown in Fig. \ref{fig:density_profiles}.
A high density profile (HDP) with a central density $\rho_0 = 10^{-12} \, \mathrm{g \, cm^{-3}}$ and a core radius $a=1.6 \times 10^{-3} \, \mathrm{pc}$ and a low density profile (LDP) with a central density $\rho_0 = 10^{-18} \, \mathrm{g \, cm^{-3}}$ and a core radius $a=2 \, \mathrm{pc}$.
Both density profiles yield a total baryonic mass $M_{gas} \sim 10^7 \, \mathrm{\Msun}$ over the entire spatial domain. In the HDP case the spatial domain has been slightly enlarged, with respect to the LDP one, to fulfill this condition.
The yellow-shaded area in Fig. \ref{fig:density_profiles} shows that the LDP may be interpreted as the density profile resulting after the formation of a black hole of mass $\sim 10^5 \, \mathrm{\Msun}$ at the center of the halo (see \citealt{Latif_2013c, Latif_2014b}).
Consequently, the HDP may be interpreted as the density profile resulting after the formation of a very small ($\lsim 10^3 \, \mathrm{\Msun}$) black hole seed, which leaves the matter distribution of the halo almost unaltered. In summary, we study four models: HDP with radiatively efficient accretion, HDP with radiatively inefficient accretion, LDP with radiatively efficient accretion, and LDP with radiatively inefficient accretion.

\section{Analytical insights}
\label{sec:results}
In this Section we present our analytic model to describe the growth of high-redshift black holes.
We predict whether the mass inflow is feeding-dominated or feedback-limited, as a function of the main physical properties of the (black hole + host halo) system. This analytic framework is then employed to determine a time-evolving spatial scale outside of which the radiative feedback is highly effective and outflows dominate. Moreover, we envisage the existence of a mass scale above which the accretion flow is always feeding-dominated.

\subsection{Modeling the mass inflow}
Our analytic model takes into account the following properties: (i) the density of the gas in the inner sections of the halo $\rho_0$, (ii) the mass of the black hole seed $M_{\bullet}$, (iii) the matter-radiation conversion factor $\epsilon$, (iv) the Eddington factor $f_{Edd}$.
Starting from quasi-static conditions, the accretion rate increases, due to the black hole's gravitational pull. The build-up of the accretion rate causes an increasing emission of radiation, which eventually might be able to stop the gas inflow, or even invert its velocity: this depends on the intensity of the radiation field and on the inertia of the gas.
The rate at which the linear momentum of the gas is changed relies on the comparison between two time scales: (i) the feedback time scale, $t_{fb}$, defined as the time needed by the radiation pressure to significantly (i.e. by a factor $e$) change $\dot{M}_{\bullet}$, and (ii) the accretion time scale, $t_{acc}$, which estimates the time needed to consume the gas mass inside the inner regions of the accretion flow.

Given the general expression for the acceleration caused by the radiation pressure (Eq. \ref{a_rad}) and the relations between the luminosity $L$ and the Eddington factor $f_{Edd}$ for radiatively efficient and inefficient flows (Eqs. \ref{radiatively_efficient}, \ref{radiatively_inefficient_1}, \ref{radiatively_inefficient_2}), we calculate $t_{fb}$.
The impulse theorem (radial component) states that $dp = F dt$, where $p$ is the linear momentum per unit volume of the gas, $F$ is the force applied on the gas per unit volume and $t$ is the time:
\begin{equation}
d(\rho v) = \rho a_{rad} dt = \rho \frac{\kappa}{4 \pi c} \frac{L}{r^2} dt \, .
\end{equation}
This equation needs to be evaluated at some radius representative of the accretion process onto the black hole. In a simulation, this radius is the innermost cell of the spatial grid, where the gas is assumed to be accreted by the black hole. From a purely theoretical point of view, it may be considered as the radius where the accretion disk forms. In either case, we designate this spatial scale $r_0$, the accretion radius.
Evaluating the previous equation at $r=r_0$ gives:
\begin{equation}
d(4 \pi r_0^2 \rho_0 v_0) = \frac{\rho_0 \kappa_0}{c} \, L_0 \,dt \, .
\end{equation}
Here, $L_0$ is computed with Eq. \ref{L_definition} and the mass flux is computed at the accretion boundary, corresponding to the accretion rate $\dot{M_{\bullet}}$.
Therefore:
\begin{equation}
d\dot{M}_{\bullet} = \frac{\rho_0 \kappa_0}{c} \, L_0 \,dt \, .
\end{equation}
We solve this ordinary differential equation with the initial condition $\dot{M}_{\bullet}(t=0)$, assuming that the accretion rate decreases with time: in the case of an increasing mass flux, only the sign of the exponential solution changes.
The general solution is:
\begin{equation}
\dot{M}_{\bullet}(t) = \dot{M}_{\bullet}(t=0) e^{-t/t_{fb}} \, ,
\label{general_solution_ODE}
\end{equation}
where
\begin{equation}
t_{fb} = \frac{\psi}{\epsilon \rho_0 \kappa_0 c} \, ,
\label{t_fb}
\end{equation}
and
\begin{equation}
\psi = f_{Edd} \frac{L_{Edd}}{L} \, .
\label{psi_definition}
\end{equation}
In the standard radiatively efficient scenario $\epsilon = 0.1$ and $\psi =1$, while in the radiatively inefficient case $\epsilon = 0.04$ and $\psi =25$ (for $f_{Edd} < 50$) and $\epsilon < 0.04$ and $\psi > 25$ (for $f_{Edd} \geq 50$).
The feedback time scale is not related to the black hole mass $M_{\bullet}$, but only to the mass of the inflowing gas, via $\rho_0$.
In the slim disk case, for a given density $\rho_0$, the feedback time scale is $\gsim 60$ times longer: the system reacts to a modification of the accretion rate in a much slower way.

To calculate $t_{acc}$, we estimate the time scale for the consumption (due to accretion) of the gas mass inside some radius $r$, given an accretion rate $\dot{M}_{\bullet}$:
\begin{equation}
t_{acc} = \frac{M_g(<r)}{\dot{M}_{\bullet}}=\frac{\epsilon c \kappa}{4 \pi G f_{Edd}} \frac{M_g(<r)}{M_{\bullet}},
\end{equation}
where on the right-hand side we have parametrized the luminosity through the Eddington factor $f_{Edd}$. The accretion time scale\footnote{The accretion time scale is also very well approximated by a fraction $1/3$ of the crossing time for a particle at some radius $r$: $M/\dot{M}=r/(3\dot{r})$.} is inversely proportional to the black hole mass $M_{\bullet}$ and directly proportional to the gas mass within the radius $r$.  

Defining the ratio of the two time scales as ${\cal T}(r,t) \equiv {t_{fb}}/{t_{acc}}$, a transition radius, $r_T$, exists such that $t_{acc}(r_{T}) = t_{fb}(r_{T})$. 
The flow is {\it feeding-dominated} in the region where $r \ll r_T$, $t_{acc} \ll t_{fb}$, ${\cal T}(r,t) \gg 1$ and gas is easily available for the accretion. The flow is instead {\it feedback-limited} where $r \gg r_T$, $t_{acc}\gg t_{fb}$, ${\cal T}(r,t) \ll 1$ because outflows dominate over inflows and accretion proceeds in an intermittent way, or can be even halted if the gas reservoir is empty. The parameter ${\cal T}(r,t)$ allows us to determine whether an accretion flow is feeding-dominated or feedback-limited. The expressions for $r_{T}$  and ${\cal T}(r,t)$ are as follows:
\begin{equation}
r_{T} = \left[ \psi \frac{3Gf_{Edd}M_{\bullet}}{(\epsilon \rho_0 \kappa_0 c)^2} \right]^{1/3};
\label{r_t_definition}
\end{equation}
and
\begin{equation}
{\cal T}(r,t) = \psi \frac{3 G f_{Edd} M_{\bullet}}{(\epsilon \rho_0 \kappa_0 c)^2 r^3} = \left( \frac{r_T}{r} \right)^3.
\label{p_parameter}
\end{equation}
Note that the black hole mass $M_{\bullet}$ increases with time, while $\rho_0$ decreases as gas is consumed, therefore the transition radius increases with time: a larger fraction of the host halo enters the feeding-dominated region, and accretion becomes progressively easier.
Moreover, we define ${\cal T}(r_0,t) \equiv {\cal T}_0$.

Eq.~\ref{p_parameter} shows that the efficiency of an accretion flow depends on several variables that we discuss in turn. Firstly, the smaller the internal density of the halo, $\rho_0$, the longer is $t_{fb}$. This is because when the gas density is low the physical accretion rate on the black hole is small, and, at a given black hole mass, radiation pressure is less effective. It will be easier to have a feeding-dominated flow in the LDP than in the HDP case. Secondly, a smaller radiative efficiency $\epsilon$ yields a less effective radiation pressure, and consequently radiatively inefficient accretion would be more likely feeding-dominated. Thirdly, a larger Eddington factor $f_{Edd} \propto \dot{M}_{\bullet} \propto v$ implies a higher inward linear momentum of the gas. In this case, the inward velocity of the gas mass is less easily inverted. Lastly, the black hole mass $M_{\bullet}$: since $\dot{M}_{Edd} \propto M_{\bullet}$, a smaller mass corresponds to a smaller physical critical accretion rate: in a given halo a small black hole is more likely to be fed at rates that, for its mass, give rise to high radiation pressure. A small black hole is therefore more likely to find itself in the feedback-limited regime.

\begin{figure}
\vspace{-1\baselineskip}
\hspace{-0.5cm}
\begin{center}
\includegraphics[angle=0,width=0.50\textwidth]{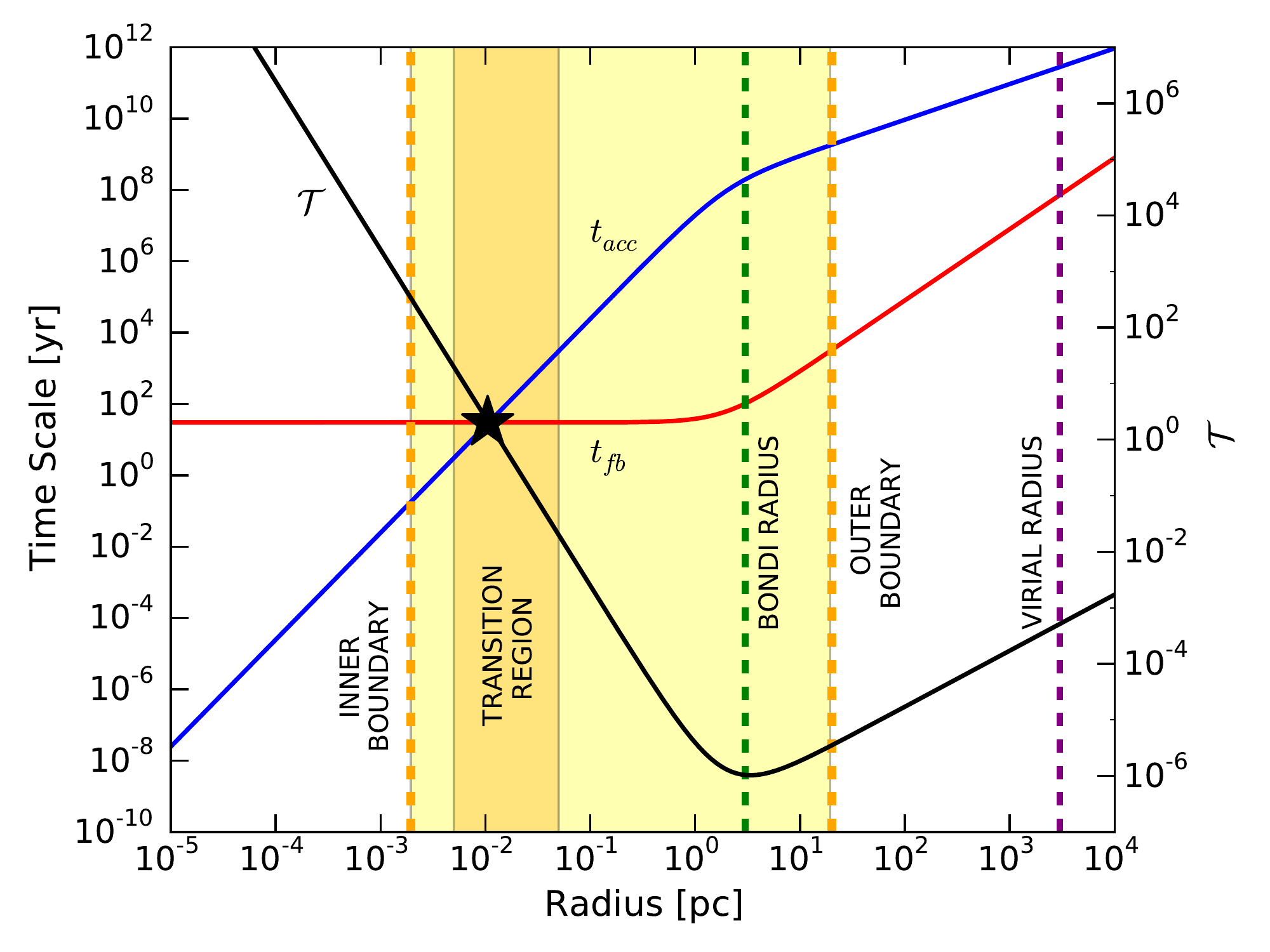}
\caption{Comparison, at $t=0$, between the accretion time scale ($t_{acc}$) and the feedback time scale ($t_{fb}$) for the accretion flow onto a black hole of initial mass $10^5 \, \mathrm{\Msun}$ in the radiatively efficient and LDP case. The black line is the value of ${\cal T}(r, t_0)$. The yellow-shaded area is the region of integration in the present work, while the orange-shaded region indicates the transition region, where $0.1 \lsim {\cal T} \lsim 10$. The inner and outer boundaries, the Bondi radius and the virial radius of the halo are shown. The black star indicates the position of the transition radius.}
\label{fig:time_scales}
\end{center}
\end{figure}

As an example, Fig. \ref{fig:time_scales} shows a comparison between the accretion time scale ($t_{acc}$) and the feedback time scale ($t_{fb}$), both computed at $t=0$, for the accretion flow onto a black hole of initial mass $10^5 \, \mathrm{\Msun}$ embedded in an LDP in the radiatively efficient case (Eq. \ref{radiatively_efficient}).
The transition radius $r_T \sim 2\times 10^{-2} \, \mathrm{pc}$ is larger than the accretion radius ($r_0 \sim 2 \times 10^{-3} \, \mathrm{pc}$), so we expect the black hole growth to be feeding-dominated. Moreover, an overall modification of the spatial velocity profile of the accretion flow should be visible around the position of the transition radius, in the spatial range that we schematically call transition region, where $0.1 \lsim {\cal T} \lsim 10$. These effects are discussed in Section~4, dedicated to the results of our numerical simulations. 

In the HDP case the transition radius would be smaller than the accretion radius and the growth would be feedback-limited. On the contrary, in the radiatively inefficient cases, both HDP and LDP, the transition radius would be even larger than the one shown in Fig. \ref{fig:time_scales}, leading to a more extended feeding-dominated region.

\subsection{Relation between ${\cal D}$ and ${\cal T}_0$}
In the following we derive an analytic relation between our model, through the quantity ${\cal T}_0$ (Eq. \ref{p_parameter}), and ${\cal D}$, the duty cycle for the black hole growth (Eq. \ref{DC_def}), which is a phenomenological way, computable only \textit{a posteriori}, to describe if the gas inflow is continuous.
Even Adaptive Mesh Refinement cosmological simulations cannot resolve the typical spatial scale of accretion, therefore they have to resort to some kind of sub-grid prescriptions to model the black hole growth, like assuming that they \textit{continuously} accrete at the Bondi-Hoyle-Lyttleton rate \citep{Bondi_1952}, capped at the Eddington rate (see e.g. \citealt{Springel_2005, DiMatteo_2008, 2013MNRAS.428.2885D, Costa_2014, 2015arXiv150400018D}). Our model may provide in such cases more realistic values for the duty cycles of these sources.

Calling $t_{idle}$ the fraction of $t_{tot}$ during which the black hole is not accreting, it is possible to show that this relation holds:
\begin{equation}
\frac{t_{idle}}{t_{tot}} = e^{-t_{fb}/2t_{acc}} \, ,
\end{equation}
by solving the following differential equation (where $\chi$ is the time scale for the variation of $\dot{M}_{\bullet}$):
\begin{equation}
-\frac{d \chi}{dt} = \frac{\chi}{2t_{acc}} \, .
\end{equation} 
The factor $2$ accounts for the fact that, once the radiation pressure exerts an action on the infalling gas for some time $\Delta t$, a time $2\Delta t$ is needed to re-establish the accretion rate preceding the acceleration (see Eq. \ref{general_solution_ODE}).
We obtain:
\begin{equation}
{\cal D} = (1-e^{-{\cal T}_0/2}) \, .
\label{P_DC_relation}
\end{equation}
With our definition, ${\cal D} ({\cal T}_0=1) \sim 0.4$. A proof of the validity of Eq. \ref{P_DC_relation} is provided in Fig. \ref{fig:fit},  described in section 4.2.1.

\subsection{The black hole - host halo connection}
The transition radius, which  separates the feeding-dominated region ($r \ll r_T$) from the feedback-limited region ($r \gg r_T$), increases with time (cf. Eq.~\ref{r_t_definition}). Consequently, if the black hole growth is feeding-dominated at $r_0$ and $t=0$, it will always be so.  Asking that ${\cal T}(r_0, t_0) \gsim 1$ translates into a black hole mass above which the flow will always be feeding-dominated:
\begin{align}
M_{crit} = & \, \frac{3 \times 10^6}{\psi f_{Edd}} \left(\frac{\epsilon}{0.1}\right)^2 \left(\frac{\rho_0 }{5 \times 10^{-15} \, \mathrm{g \, cm^{-3}} }\right)^2 \times \nonumber \\ 
& \times \left(\frac{r_0 \, }{10^{-4} \, \mathrm{pc}}\right)^3 \, \mathrm{\Msun} \, .
\label{M_limit}
\end{align}
Here, we considered $r_0 \sim 10^{-4}-10^{-3} \, \mathrm{pc} \sim 20-200 \, \mathrm{AU}$ as a typical spatial scale for accretion disks.

In the HDP case for the standard accretion scenario the previous limit reads:
\begin{equation}
{\cal T}(r_0, t_0) > 1 \, \, \, \mathrm{if} \, \, \, M_{\bullet} > M_{crit} \sim 3 \times 10^6 \, \mathrm{\Msun} \, \, \, \mathrm{(HDP \, -\, Std)} \, ,
\end{equation}
while both in the LDP case and in the slim disk accretion scenario (both density profiles) the limit is negligible:
\begin{equation}
{\cal T}(r_0, t_0) > 1 \, \, \, \mathrm{if} \, \, \, M_{\bullet} > M_{crit} \sim 10-100 \, \mathrm{\Msun} \, \, \, \mathrm{(other \, cases)} \, .
\end{equation}

The physical meaning of $M_{crit}$ requires a clarification, since one normally expects that feedback halts the black hole growth above a given mass, rather than below (see e.g. \citealt{Silk_1998, King_2003, King_2010}). The meaning of the lower limit we find is that when $M_{\bullet} > M_{crit}$ the accretion rate needed to exert a sufficiently strong feedback is so high, for the halo in question, that the accretion flow cannot produce it. In other words, the inflow rate is determined by the halo properties, and is, at least initially, independent of the black hole mass. If a given halo provides the same $\dot{M}_{\bullet}$ to a small black hole or a large black hole, it will be the smaller black hole that will reach the Eddington limit first, having its growth stunted. As a general result, smaller black hole seeds should encounter great challenges during the first stages of the growth, characterized by outflows and very low values of the duty cycle. This effect could play an important role at high redshifts, where black hole seeds of different mass may form from the same host halo, depending on the thermal and radiative properties of the environment.  

\section{Numerical results}
The following two subsections describe our numerical simulations and their analysis through the growth model outlined so far.

\subsection{Inflows and outflows}
\begin{figure}
\vspace{-1\baselineskip}
\hspace{-0.5cm}
\begin{center}
\includegraphics[angle=0,width=0.50\textwidth]{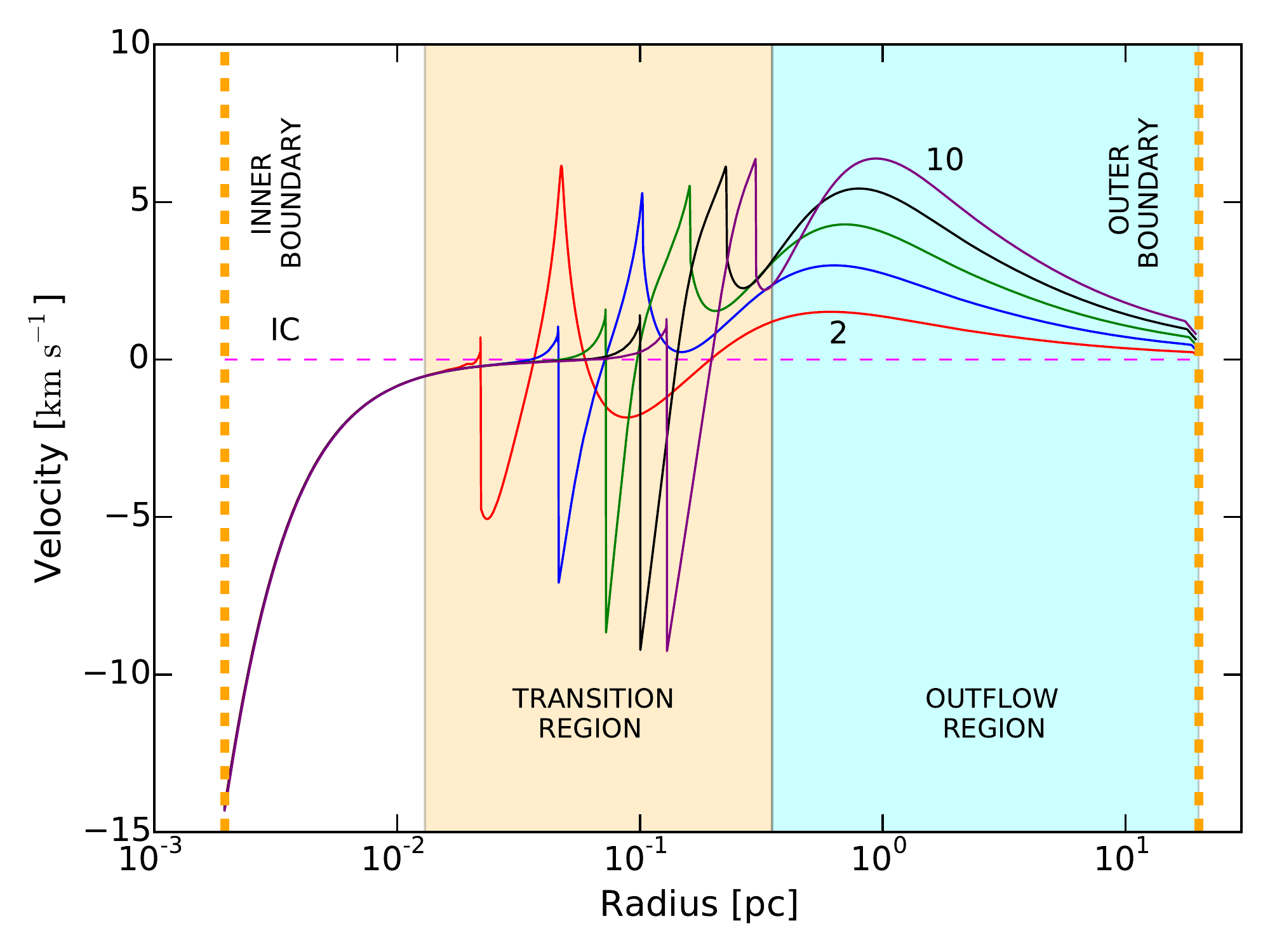}
\vspace{-0.1cm}
\includegraphics[angle=0,width=0.50\textwidth]{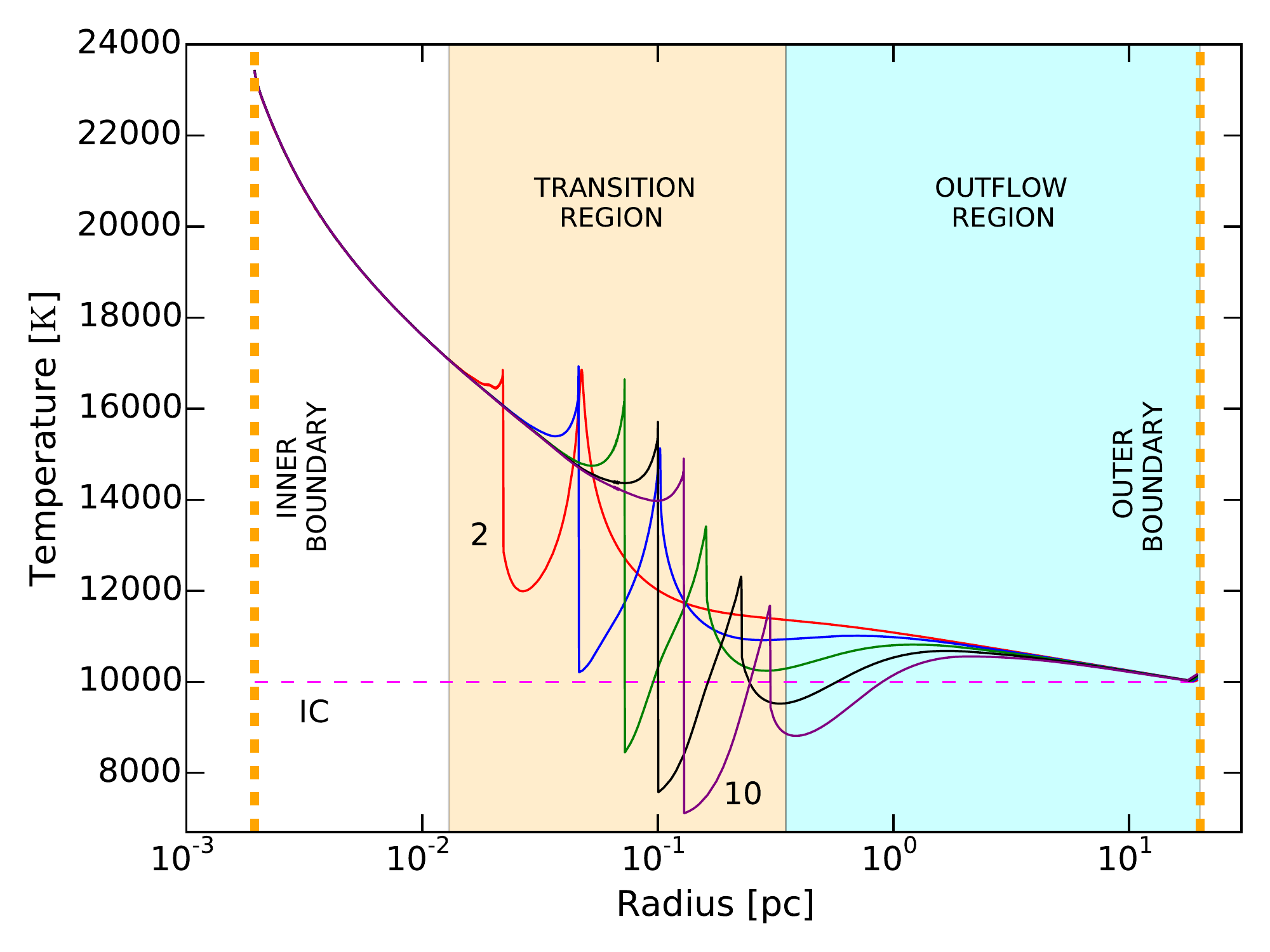}
\caption{Velocity and temperature spatial profiles for the accretion flow onto a $M_0 = 10^5 \, \mathrm{\Msun}$ black hole embedded in a LDP density profile, accreting in the radiatively efficient mode. The total integration time is $t_{tot} = 10^{5}\, \mathrm{yr}$ ($t_{tot} = i \, \Delta t$ with $\Delta t = 10^4 \, \mathrm{yr}$ and $i=0, \, 2, \, 4, \, 6, \, 8, \, 10$). The initial conditions are marked with the label ``IC". The transition and outflow regions are defined by $0.1 \lsim {\cal T}(r) \lsim 10$ and ${\cal T}(r) \lsim 0.1$ respectively. The inner and outer boundaries are shown with orange dashed lines.}
\label{fig:velocity_profile}
\end{center}
\end{figure}

As an example, we discuss a simulation of the accretion flow onto a black hole of initial mass $10^5 \, \mathrm{\Msun}$ in the radiatively efficient and LDP case.
Starting from static and isothermal conditions, the gravity of the black hole rapidly pulls in the gas, building up the accretion flow.
Fig. \ref{fig:velocity_profile} shows the time evolution of the velocity and temperature spatial profiles, which may be interpreted as the dynamical and thermal counterparts of Fig. \ref{fig:time_scales}, bearing in mind that $r_T(t)$ always increases.
The transition and the outflow regions are defined as the set of radii $r$ where, at least once during the time evolution of the system, the relations $0.1 \lsim {\cal T}(r) \lsim 10$ and ${\cal T}(r) \lsim 0.1$ respectively hold.
In the region close to the accretion boundary the inflow is very smooth, with inward velocities up to $\sim 15 \, \mathrm{km \, s^{-1}}$ and temperatures rising up to $\sim 2.4 \times 10^4 \, \mathrm{K}$.
In the transition region, the flow starts to be disturbed by radiative feedback, which becomes more effective due to the increase of $t_{acc}$, with frequent velocity inversions and a more complex temperature profile.
In the outflow region the radiative feedback is dominant, with large (up to $\sim 5 \, \mathrm{km \, s^{-1}}$) outflowing velocities and a temperature profile which reconnects to the thermal floor ($T \sim 10^4 \, \mathrm{K}$) of the host halo.

\begin{figure}
\vspace{-1\baselineskip}
\hspace{-0.5cm}
\begin{center}
\includegraphics[angle=0,width=0.50\textwidth]{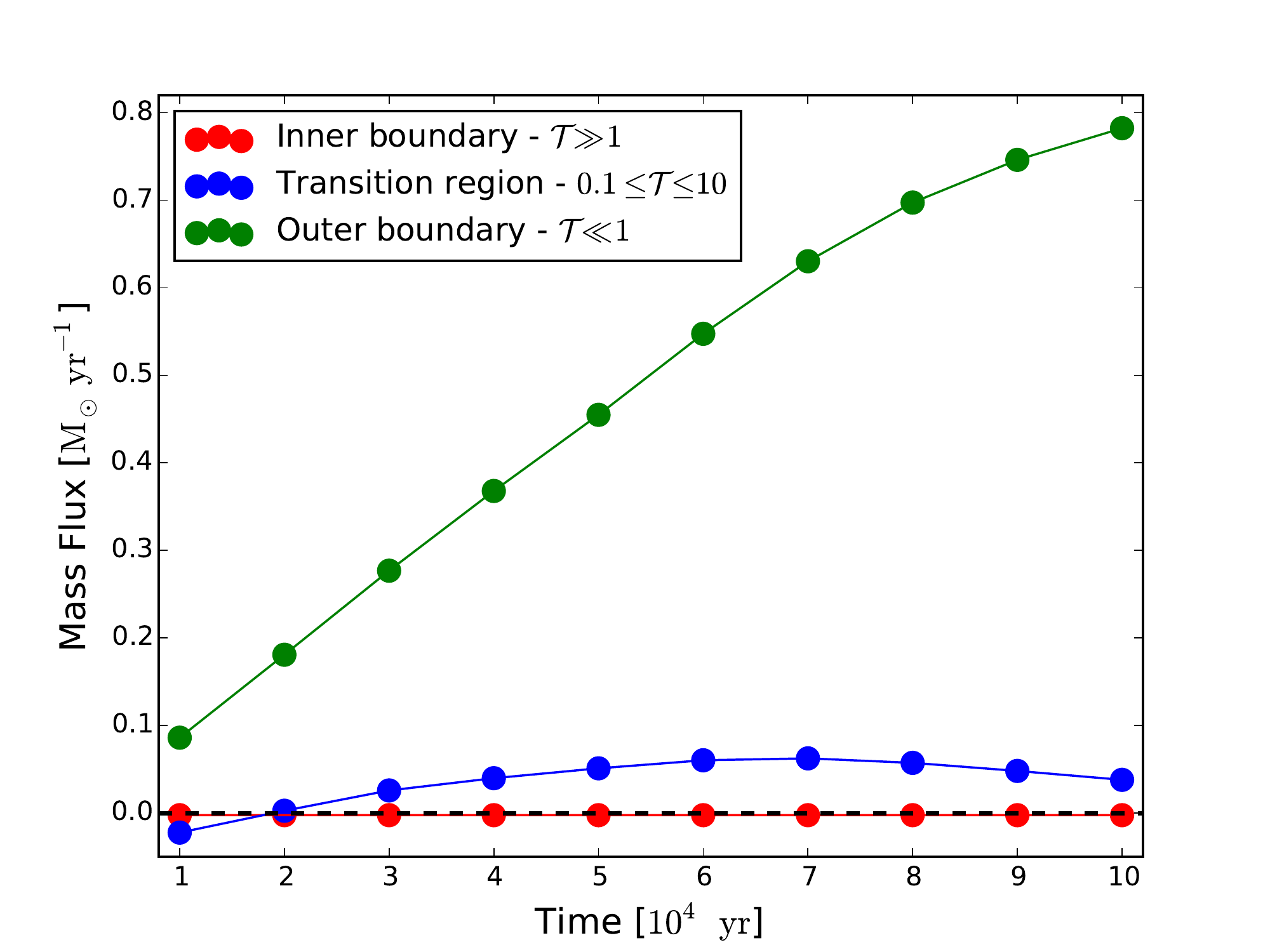}
\caption{Time evolution of the mass flux computed at the inner boundary, outer boundary and inside the transition region highlighted in Fig. \ref{fig:velocity_profile}. There is a slow but constant ($\sim 2.0 \times 10^{-3} \, \mathrm{\Msun \, yr^{-1}}$) accretion at the inner boundary, while the outer region is characterized by a strong outflow, with a mass flux up to $\sim 0.8 \, \mathrm{\Msun \, yr^{-1}}$. The transition region is characterized by a mild outflow, on average.}
\label{fig:mass_flux}
\end{center}
\end{figure}

For the same simulation, Fig. \ref{fig:mass_flux} shows the time evolution of the mass flux at the inner boundary (i.e. the accretion rate), at the outer boundary and inside the transition region, in which the computed values are a spatial average over the cells belonging to this region at each time of the simulation.
The inner region is characterized by a slow, but \textit{constant}, accretion of order $\sim 2.0 \times 10^{-3} \, \mathrm{\Msun \, yr^{-1}}$. The outer region is swept by large outflows, whose magnitude increases with time, reaching a peak of $\sim 0.8 \, \mathrm{\Msun \, yr^{-1}}$, while the transition region is characterized by a mild outflow.

In the HDP case (for the same accretion scenario), $r_T \lsim r_0$ and the growth is feedback-limited: the accretion is discontinuous and the transition region would extend down to the accretion boundary, while the outflows would be more intense. On the contrary, in the radiatively inefficient cases, both HDP and LDP, $r_T \gg r_0$: the accretion flow would be smooth and continuous over a large fraction of the spatial domain, with the transition region starting to be visible only close to the outer radius.

\subsection{The growth efficiency}
In the following we numerically test the predictions of our model with 16 runs, in order to adequately explore the range of density profiles (HDP and LDP), black hole mass seeds ($10^{3-6} \, \mathrm{\Msun}$) and radiative efficiencies ($L \propto f_{Edd}$ and $L \propto \ln f_{Edd}$).
The simulations are run for $\sim 10^5 \, \mathrm{yr}$ to allow all the simulated accretion flows to reach steady-state conditions without arriving at a complete depletion of the gas reservoir. The parameters reported in the figures, $\langle{\cal T}(r_0)\rangle$, $\langle f_{Edd} \rangle$ and $\langle{\cal D}\rangle$, are an average over the entire simulation time, while the parameter $\Delta M/M_0$ expresses the total mass growth at the end of the simulation, normalized to the initial value of the seed mass. Fig. \ref{fig:comparison_standard} shows the results in the radiatively efficient case, while Fig. \ref{fig:comparison_SD} refers to the radiatively inefficient one. The top panels refer to HDP, while the bottom ones to LDP.

\subsubsection{The $M_{\bullet}=10^{3-6} \, \mathrm{\Msun}$ runs}
\begin{figure}
	\includegraphics[angle=0,width=0.50\textwidth]{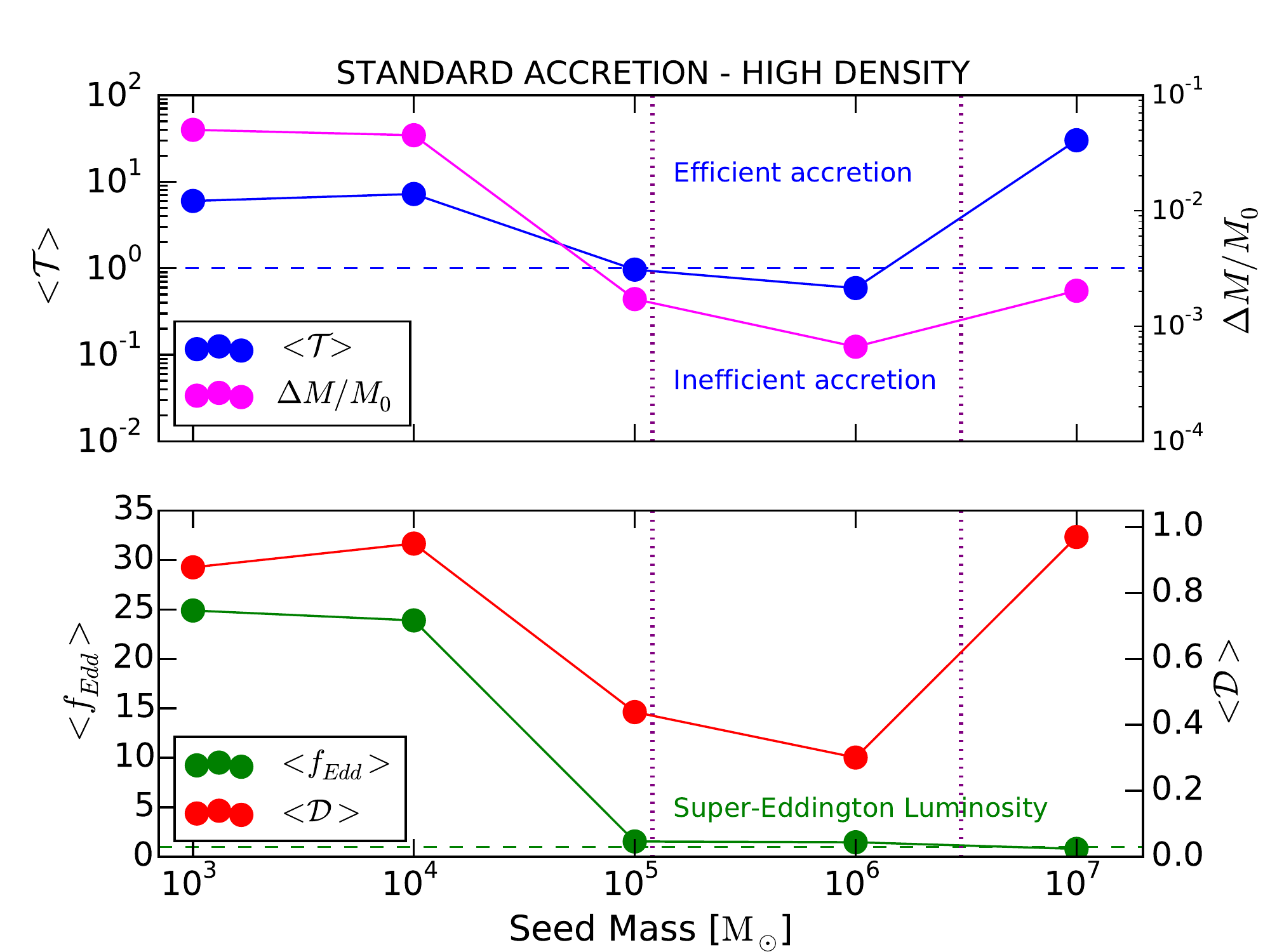}
	\vspace{-0.1cm}
	\includegraphics[angle=0,width=0.50\textwidth]{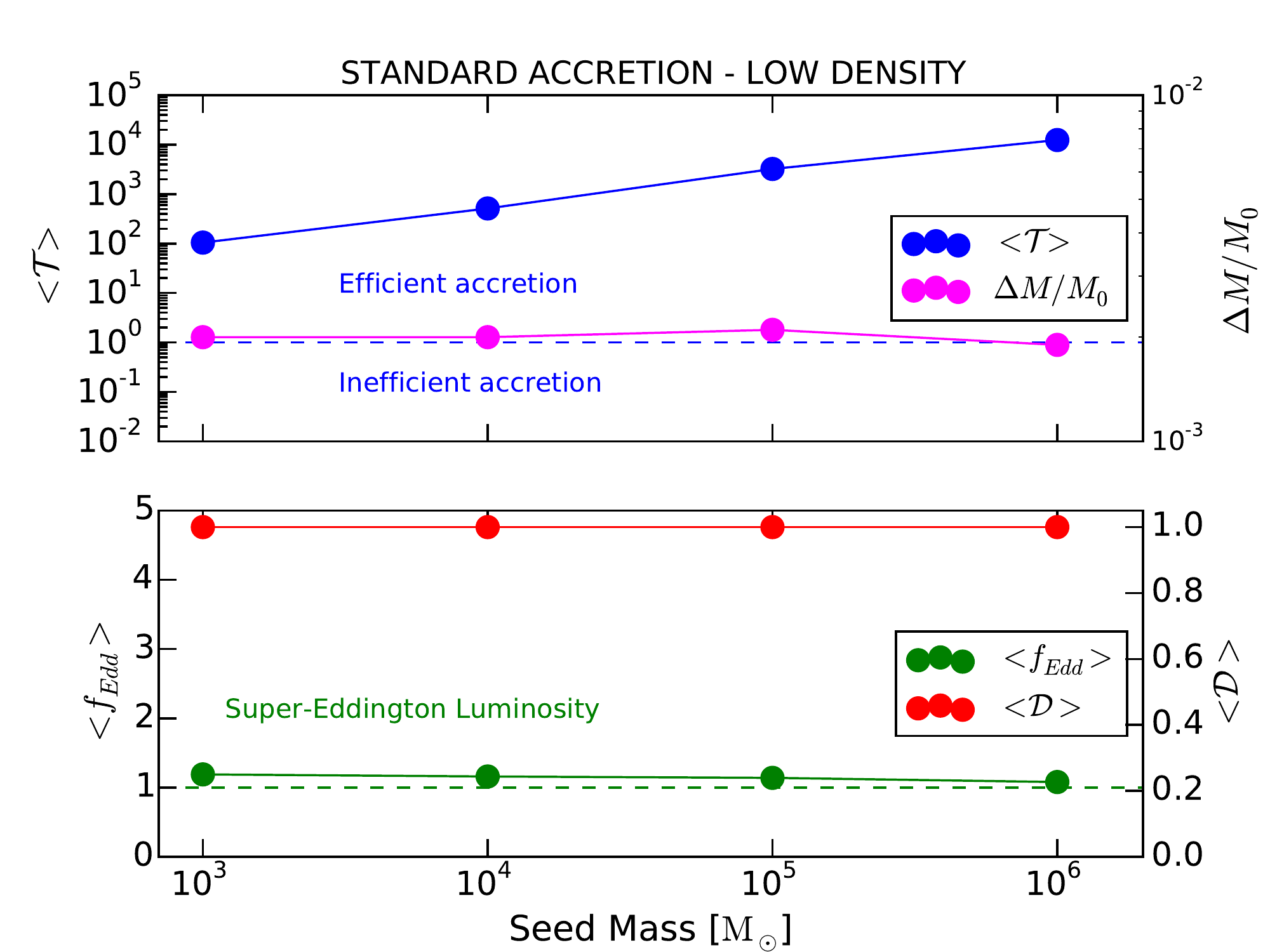}
	\vspace{-0.1cm}
	\caption{Radiatively efficient case - Comparison between the parameters used to describe the accretion flow (${\cal T}$, $\Delta M/M_0$, $f_{Edd}$ and ${\cal D}$), for black hole seed masses in the range $10^{3-6} \, \mathrm{\Msun}$ and two density profiles (HDP above and LDP below). The vertical purple lines show the values of $M_{crit}$ for $f_{Edd}=25$ and $f_{Edd}=1$, from left to right. The physical accretion rates are, in ascending order of mass: $4.0 \times 10^{-4} \, \mathrm{\Msun \, yr^{-1}}$, $4.5 \times 10^{-3} \, \mathrm{\Msun \, yr^{-1}}$, $1.5 \times 10^{-3} \, \mathrm{\Msun \, yr^{-1}}$, $6.5 \times 10^{-3} \, \mathrm{\Msun \, yr^{-1}}$ for the HDP, and $2.0 \times 10^{-5} \, \mathrm{\Msun \, yr^{-1}}$, $2.0 \times 10^{-4} \, \mathrm{\Msun \, yr^{-1}}$, $2.0 \times 10^{-3} \, \mathrm{\Msun \, yr^{-1}}$, $2.0 \times 10^{-2} \, \mathrm{\Msun \, yr^{-1}}$ for the LDP. The simulation with $M_0 = 10^7 \, \mathrm{	\Msun}$ shows a very large value for the physical accretion rate, $\sim 2.0 \times 10^{-1} \, \mathrm{\Msun \, yr^{-1}}$. See the main text for further details.}
	\label{fig:comparison_standard}
\end{figure}

The HDP case in the radiatively efficient scenario (Fig. \ref{fig:comparison_standard}, top panel) is particularly interesting because ${\cal T}$ crosses unity twice, allowing us to test the ${\cal D}-{\cal T}$ relation.  The simulated masses are all below $M_{crit} \sim 3 \times 10^6$ characteristic of this density profile and radiative efficiency for $f_{Edd}=1$, therefore one expects the flow not be feeding-dominated at all times. However, \citealt{Pacucci_2015} show that super-critical accretion is feasible, for short times, also in radiatively efficient scenarios. Indeed, in the mass range $10^{3-4} \, \mathrm{\Msun}$ the entire galaxy is in outflow ($r_T \sim r_0$), but the radiation pressure is active only when the gas accretion is underway, while during the remaining time the inflow builds up again. As a consequence, the radiation pressure is unable to sweep {\it all} the gas away from the accretion boundary, and a small physical accretion rate is sufficient to grow such small black holes, so the flow can sustain, on average, $f_{Edd} \sim 25$, with duty cycles ${\cal D} \sim 0.8-0.9$. The relevant $M_{crit}$, therefore, is not the one for  $f_{Edd} \sim 1$, but for $f_{Edd} \sim 25$. The values of ${\cal T}$ and $r_T$ slowly increase as the initial mass of the seed $M_{\bullet}$ approaches $M_{crit}\sim 1.2\times 10^5 \, \mathrm{\Msun}$ for $f_{Edd} \sim 25$ (see the left-most vertical line in Fig. \ref{fig:comparison_standard}, top panel).
 
In the range $10^{5-6} \, \mathrm{\Msun}$ a super-Eddington flow is no more sustainable, since the radiation pressure is progressively more powerful: the flow stabilizes at $f_{Edd} \sim 1$. For an Eddington-limited flow, $M_{crit} \sim 3\times 10^6 \, \mathrm{\Msun}$ (see the right-most vertical line in Fig. \ref{fig:comparison_standard}, top panel). In the process of going from $f_{Edd}\sim 25$ to $f_{Edd}\sim 1$ the flow becomes mildly feedback-limited (${\cal T} \sim 0.7$ for $M_{\bullet}=10^6 \, \mathrm{\Msun}$, ${\cal D} \sim 0.3-0.5$).
The physical accretion rates in the HDP case are within the range $10^{-3} - 10^{-4} \, \mathrm{\Msun \, yr^{-1}}$ (see Fig. \ref{fig:comparison_standard} for further details).

In the LDP case of the radiatively efficient scenario, the critical mass value $M_{crit}$ is $<10^3 \, \mathrm{\Msun}$: as a consequence the growth is feeding-dominated (${\cal D} \sim 1$ and ${\cal T} \gsim 10^2$) for all runs and accretion rates are stable, close to the Eddington value ($f_{Edd} \sim 1$).
The physical accretion rates are within the range $10^{-2} - 10^{-5} \, \mathrm{\Msun \, yr^{-1}}$, increasing with the mass $M_{\bullet}$ just as the Eddington rate.

\begin{figure}
	\includegraphics[angle=0,width=0.50\textwidth]{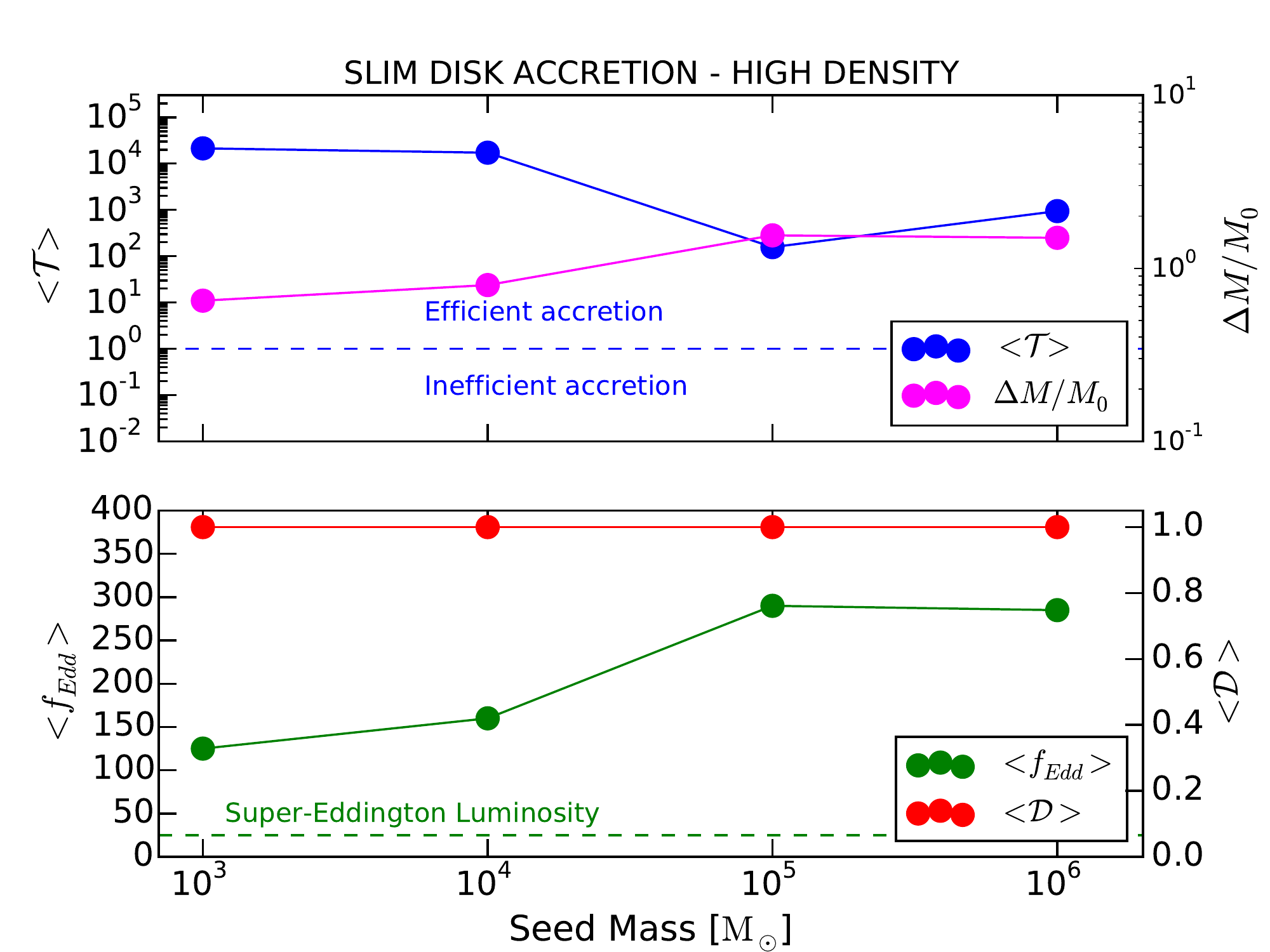}
	\vspace{-0.1cm}
	\includegraphics[angle=0,width=0.50\textwidth]{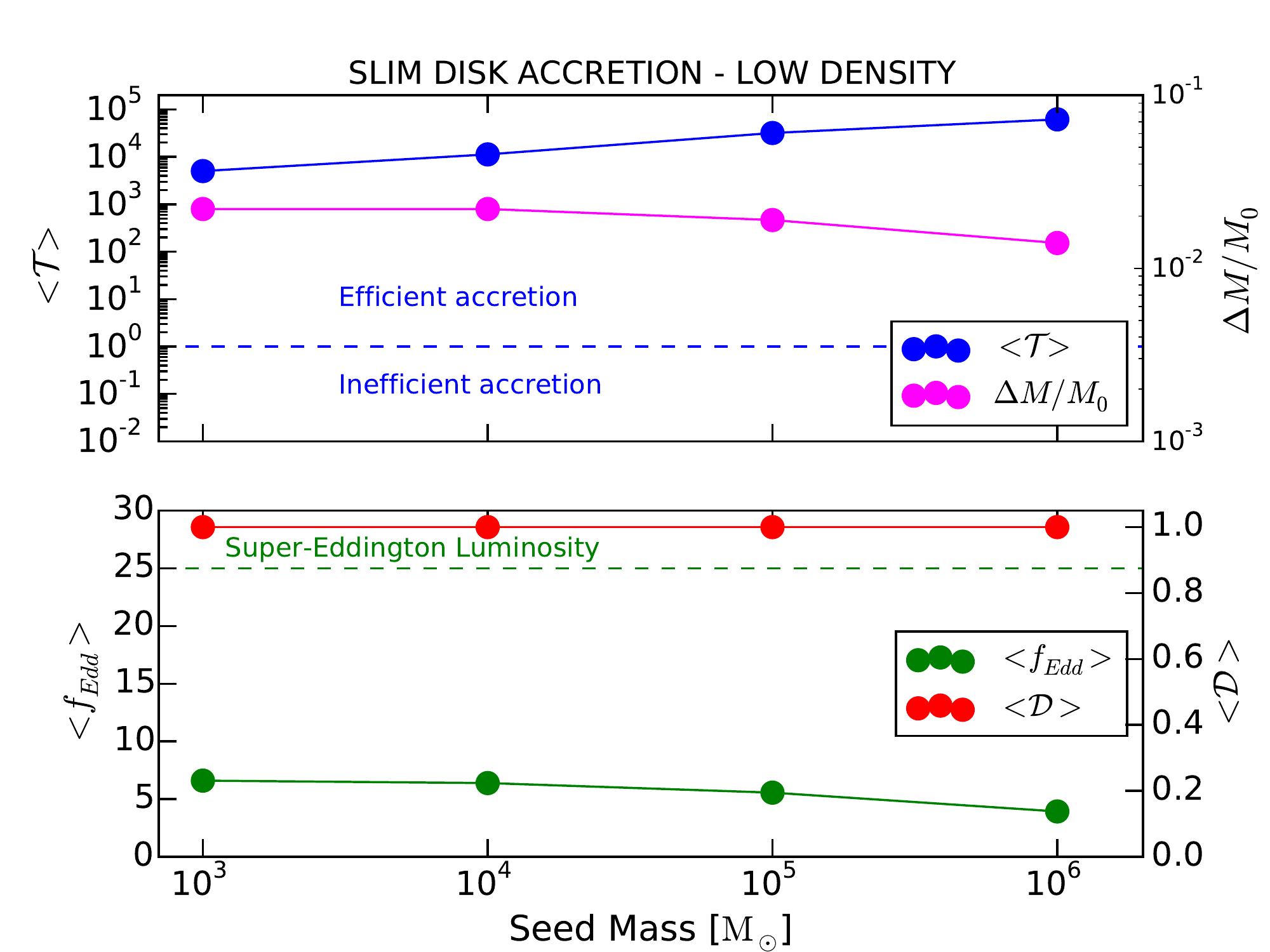}
	\vspace{-0.1cm}
	\caption{Radiatively inefficient case (slim disk) - Same as Fig. \ref{fig:comparison_standard}. The physical accretion rates are, in ascending order of mass: $7.0 \times 10^{-3} \, \mathrm{\Msun \, yr^{-1}}$, $7.0 \times 10^{-2} \, \mathrm{\Msun \, yr^{-1}}$, $1.5 \, \mathrm{\Msun \, yr^{-1}}$, $1.4 \times 10^{1} \, \mathrm{\Msun \, yr^{-1}}$ for the HDP, and $2.0 \times 10^{-4} \, \mathrm{\Msun \, yr^{-1}}$, $2.0 \times 10^{-3} \, \mathrm{\Msun \, yr^{-1}}$, $2.0 \times 10^{-2} \, \mathrm{\Msun \, yr^{-1}}$, $1.5 \times 10^{-1} \, \mathrm{\Msun \, yr^{-1}}$ for the LDP. See the main text for further details.}
	\label{fig:comparison_SD}
\end{figure}

In the HDP case of the radiatively inefficient scenario (Fig. \ref{fig:comparison_SD}, top panel) the situation is similar, because $M_{crit}<10^3 \, \mathrm{\Msun}$: so $r_T \gg r_0$ and the growth is always feeding-dominated (${\cal T}$ up to $\sim 10^4$). Accretion is not capped at the Eddington rate, so we reach rates as high as $f_{Edd} \sim 300$, leading to a \textit{super-Eddington} emitted luminosity $L \sim 5L_{Edd}$ (Eq. \ref{radiatively_inefficient_2}).
The physical accretion rates reach large values, with a peak of $14 \, \mathrm{\Msun \, yr^{-1}}$.

Similarly, in the LDP case of the radiatively inefficient scenario (Fig. \ref{fig:comparison_SD}, bottom panel) all accretion flows are feeding-dominated, with physical accretion rates within the range $10^{-1} - 10^{-4} \, \mathrm{\Msun \, yr^{-1}}$.
While the Eddington factor is $f_{Edd} \sim 5$ for all masses, the emitted luminosity is strongly \textit{sub-Eddington}, due to the low mass density of the LDP scenario, which is unable to produce a mass inflow leading to super-critical luminosities.

In summary, radiatively inefficient accretion allows for largely feeding-dominated growths of the central black hole, while standard accretion scenarios may be feedback-limited when the gas density is very high. Moreover, the main difference between the two density profiles, notwithstanding the accretion mode, is related to the accretion rates that they can sustain to feed the black hole: the HDP produces an accretion rate up to $f_{Edd} \sim 300$, decreasing dramatically the growth time of the central object.

\begin{figure}
\vspace{-1\baselineskip}
\hspace{-0.5cm}
\begin{center}
\includegraphics[angle=0,width=0.50\textwidth]{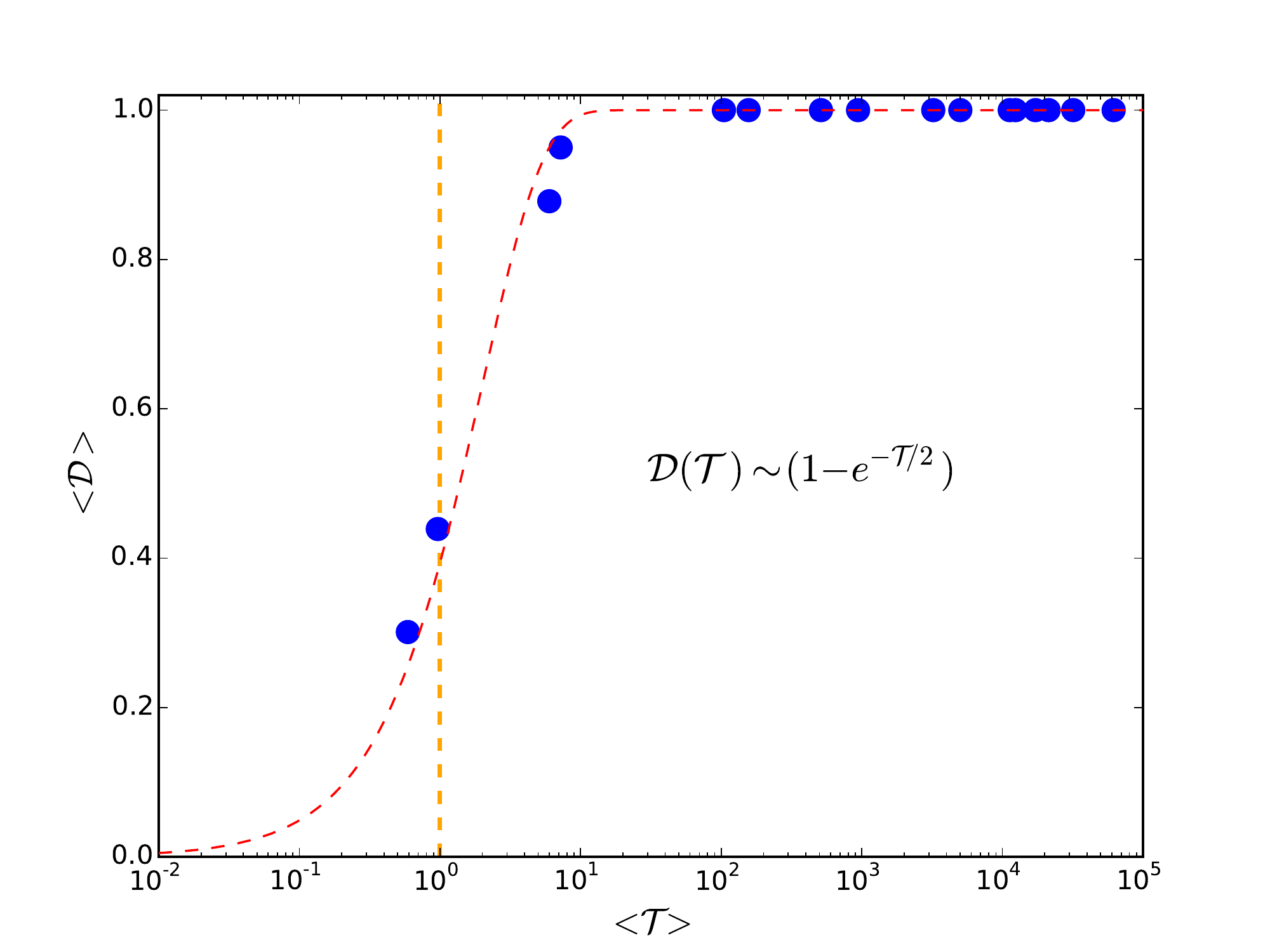}
\caption{Data points in the $({\cal D},{\cal T})$ plane, obtained from the 16 runs performed (Figs. \ref{fig:comparison_standard} and \ref{fig:comparison_SD}). The overplotted line is the theoretical relation between ${\cal D}$ and ${\cal T}$, discussed in the main text (Eq. \ref{P_DC_relation}).}
\label{fig:fit}
\end{center}
\end{figure}

With the average values of ${\cal T}$ and ${\cal D}$ computed so far for the 16 runs, we are in the conditions of testing our model for the ${\cal D}-{\cal T}$ relation, given by Eq. \ref{P_DC_relation}.
Fig. \ref{fig:fit} shows that our analytic model offers a very good fit to the points in the $({\cal D},{\cal T})$ plane.

\begin{figure}
\vspace{-1\baselineskip}
\hspace{-0.5cm}
\begin{center}
\includegraphics[angle=0,width=0.50\textwidth]{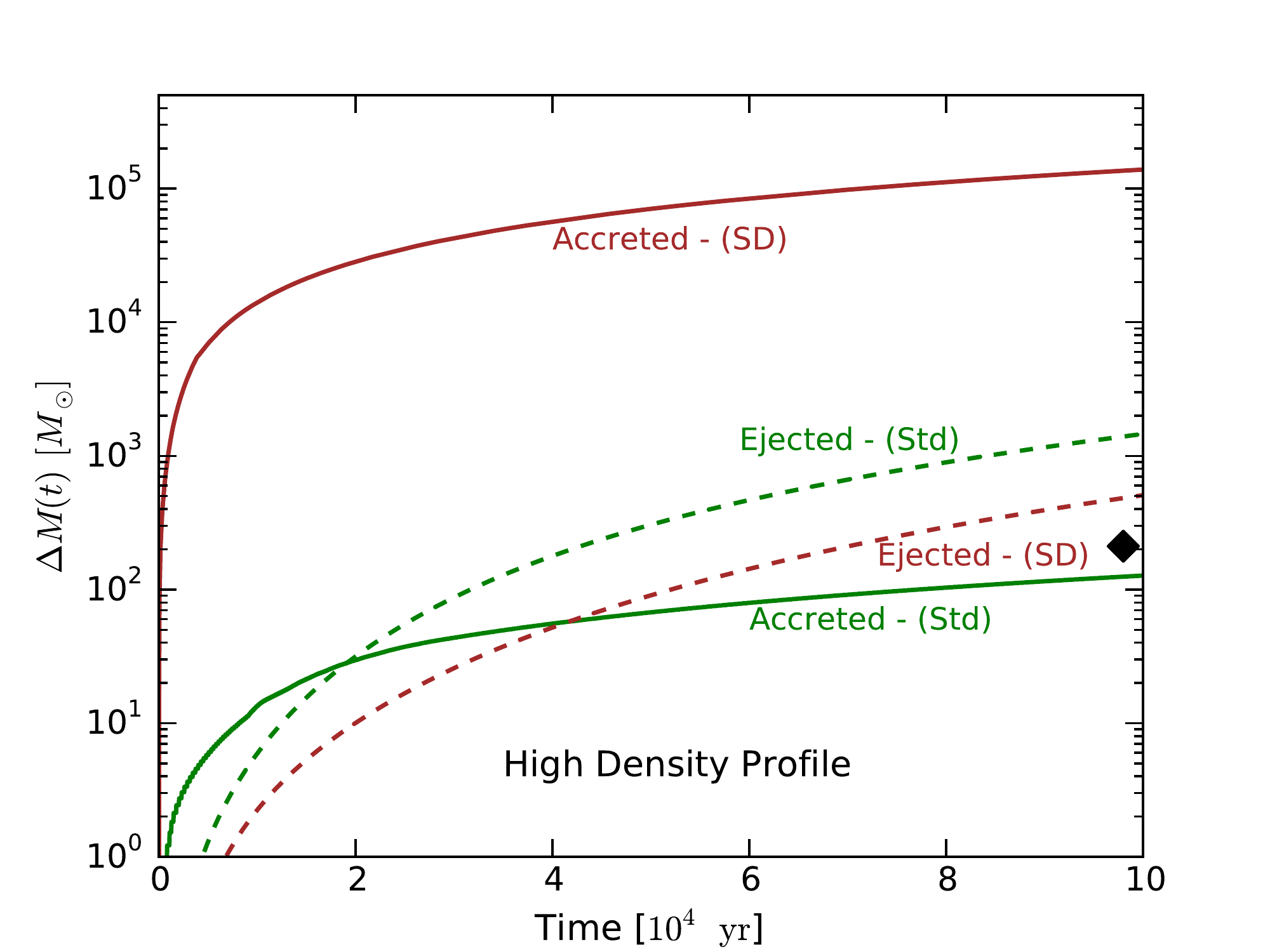}
\vspace{-0.1cm}
\includegraphics[angle=0,width=0.50\textwidth]{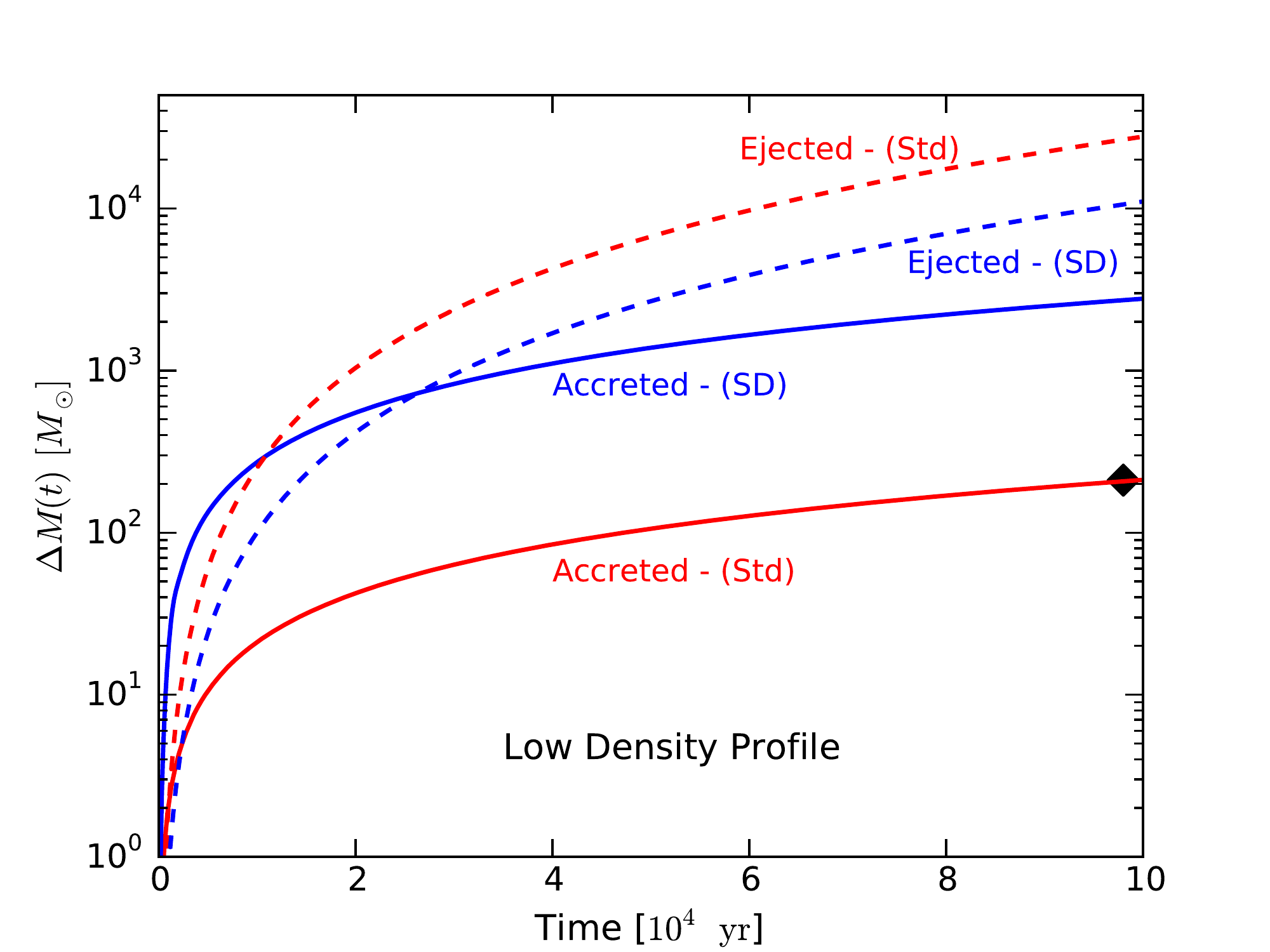}
\caption{Mass accreted (solid lines) and ejected (dashed lines) for a black hole seed with initial mass $10^5 \, \mathrm{\Msun}$, embedded in a HDP (top panel) and in a LDP (bottom panel), in radiatively efficient and inefficient scenarios. The black solid diamonds indicate the final mass that the same seed would reach accreting continuously at the Eddington rate with $\epsilon = 0.1$.}
\label{fig:mass_trend}
\end{center}
\end{figure}

In order to have a general overview of the mass accreted (across the inner boundary) and ejected (across the outer boundary) in the different simulations performed, Fig. \ref{fig:mass_trend} provides a mass balance summary.
The mass growth ($\Delta M_{\bullet} \equiv M_{\bullet}(t)-M_0$) for a black hole seed of initial mass $10^5 \, \mathrm{\Msun}$ is shown with solid lines, while the dashed lines indicate the mass ejected through outflows. The black solid diamonds indicate the final mass that the same seed would reach accreting \textit{continuously} at the Eddington rate (${\cal D} =1$, $f_{Edd}=1$) with $\epsilon = 0.1$.
The ejected mass is larger than the accreted mass, except for the HDP in the slim disk scenario (brown line), due to the large values of $f_{Edd}$ available under these physical conditions.
The mass growth in the standard radiatively efficient scenario is smaller than or equal to the one predicted by a continuous accretion at the Eddington rate, while in the slim disk scenario it is $2-3$ orders of magnitude larger (in particular, the slim disk and HDP accretion provides a final mass which is $\sim 1000$ times larger).

In a preceding work \citep{Pacucci_2015}, the authors found that only $\sim 10\%$ of the halo mass is expelled with outflows during the accretion process. In the present work, during the first $10^5 \, \mathrm{yr}$ of evolution of a $10^5 \, \mathrm{\Msun}$ seed embedded in a HDP halo the outflowing gas is $\sim 85 \%$ of the total. This difference is explained by three reasons. Firstly, the relative importance of outflows decreases with time, due to the fact that the gradient of $r_T(t)$ is positive: the simulation presented in \cite{Pacucci_2015} is $\sim 1000$ times more extended in time than the ones analyzed here. Secondly, the spatial range of the present work is much larger, then more apt to probe the outflow regions. Lastly, the halo density profiles are different, since the one employed in \cite{Pacucci_2015} had a higher central density. Most of the baryonic mass of a $T_{vir} \sim 10^4 \, \mathrm{K}$ halo at $z \sim 10$ was confined inside a sphere of radius $\sim 3 \, \mathrm{pc}$, leading to values of the optical depths of order $N_H \sim 9\times10^{25} \, \mathrm{cm^{-2}}$ at the beginning of the collapse, which may have strongly reduced the impact of outflows.

\subsubsection{The $M_{\bullet}=10^7 \, \mathrm{\Msun}$ run: a test case for $M_{crit}$}
In the radiatively efficient case with HDP (Fig. \ref{fig:comparison_standard}, top panel) the value of ${\cal T}$ reaches a minimum ($\sim 0.7$) around $M_{\bullet} \sim 10^6 \, \mathrm{\Msun}$ and rises up to $\sim 60$ again for $M_{\bullet} \sim 10^7 \, \mathrm{\Msun}$, allowing us to test the value of $M_{crit} \sim 3 \times 10^6 \, \mathrm{\Msun}$ (Eq. \ref{M_limit}) predicted for this scenario.
For $M_{\bullet} \gsim M_{crit}$, the Eddington rate is so large ($\gsim 0.3 \, \mathrm{\Msun \, yr^{-1}}$) that the accretion flow from the halo simply cannot provide it (the free-fall rate at $t=0$ is $\sim 0.2 \, \mathrm{\Msun \, yr^{-1}}$). As a consequence, the Eddington factor remains below unity ($f_{Edd} = 0.7$) while ${\cal D} \sim 1$.
In this case, the physical accretion rates are very large ($\sim 0.2 \, \mathrm{\Msun \, yr^{-1}}$), a factor $\sim 100$ higher than the values for other seed masses (see the caption of Fig. \ref{fig:comparison_standard}).
This feeding-dominated simulation has produced continuous accretion with very large rates and with sub-Eddington luminosities.

\subsection{The final black hole mass}
The integration time for all simulations is much shorter than the typical evolutionary time scale for these systems \citep{Pacucci_2015, Pacucci_F_V_2015}. In this section we estimate the depletion time $t_{end}$ needed to void the inner regions of the host halo from its gas content, due to both gas accreted by the black hole, and ejected through outflows.
Furthermore, from $t_{end}$ we can provide a rough estimate of the final black hole mass, by extrapolating the average accretion rates up to $t_{end}$.

Extrapolating the lines in Fig. \ref{fig:mass_trend}, we find the depletion time $t_{end}$ with the following condition:
\begin{equation}
M_{acc}(t=t_{end}) + M_{ej}(t=t_{end}) = M_{gas} \, ,
\end{equation}
where $M_{acc}(t)$ is the accreted mass, $M_{ej}(t)$ is the ejected mass and $M_{gas} = 10^7 \, \mathrm{\Msun}$ is the total gas mass within our computational domain.
Table \ref{tab:outline} provides a general outline of the accretion history for a black hole seed with initial mass $M_{\bullet} = 10^{5} \, \mathrm{\Msun}$ in the four scenarios investigated so far, including the depletion time $t_{end}$, the extrapolated final mass of the black hole $M_{\bullet}(t_{end})$ and its ratio with the initial baryonic mass of the host halo $M_{\bullet}(t_{end})/M_{gas}$.

\begin{table*}
\begin{minipage}{170mm}
\begin{center}
\caption{Analytic estimate of the accretion history for a black hole seed with initial mass $M_{\bullet} = 10^{5} \, \mathrm{\Msun}$ in the four scenarios investigated so far.}
\label{tab:outline}
\begin{tabular}{|c|c|c|c|}
\hline\hline
Accretion scenario & $t_{end} \, \mathrm{[Myr]}$ & $M_{\bullet}(t_{end})\, \mathrm{[\Msun]}$ & $M_{\bullet}(t_{end})/M_{gas}$\\
\hline
Standard accretion - HDP & $225 $    & $7.4\times10^{5}$  & $7\%$  \\
Standard accretion - LDP  & $ 110 $    &$1.5\times10^{6}$  & $15\%$ \\
Slim Disk accretion - HDP & $14$    & $1.0 \times 10^{7} $ & $100\% $\\
Slim Disk accretion - LDP & $9$    & $8.3 \times 10^{6} $ & $83\% $\\
\end{tabular}
\end{center}
\end{minipage}
\end{table*} 
While in standard accretion scenario the typical time scale is $\sim 100 \, \mathrm{Myr}$ \citep{Pacucci_2015} and the black hole can accrete $\sim 5\%-15 \%$ of the baryonic mass of the host halo, in radiatively inefficient modes the growth is much more rapid and efficient. Specifically, in the slim disk scenario we predict that the evolutionary time scale is of order $\sim 10 \, \mathrm{Myr}$ \citep{Pacucci_F_V_2015}, with outflows playing a negligible role: the black hole is able to accrete $\sim 80\%-100\%$ of the host halo gas.

\subsection{A bimodal evolution of the black hole seeds}
Several works in literature (e.g. \citealt{Silk_1998, King_2003, King_2010}) have investigated various forms of the so-called $M_{\bullet}-\sigma$ relation, which provides an upper limit for the black hole mass: a compact object embedded in a halo with velocity dispersion $\sigma \sim v_{esc}$ (the halo escape speed) can grow up to a mass given by the $M_{\bullet}-\sigma$ relation, while the remaining gas is dispersed by radiation-driven \citep{Silk_1998} and/or momentum-driven \citep{King_2003} outflows. 
The usual assumption adopted in these works is that the momentum flux ($\dot{M}_{out} v$) of the outflowing gas is comparable to the one in the Eddington-limited radiation field: $\dot{M}_{out} v \sim L_{Edd}/c$.
In this work we have taken an alternative view where, comparing the time scales for gas infall and gas ejection as a function of radius, we prove that the momentum flux may be very different from the Eddington value: for instance, in the slim disk model where super-critical accretion rates may be associated with sub-Eddington luminosities. 

Our approach predicts the existence of a critical black hole mass $M_{crit}$ \textit{above} which the accretion is negligibly affected by outflows: this, in turn, may lead to a bimodal evolution of the initial mass function of high-redshift black hole seeds. The lower-mass seeds ($M_{\bullet}<M_{crit}$) would go through a feedback-limited growth, with recurring episodes of strong outflows which deplete the inner regions of the host halo from its mass content: the black hole cannot accrete more than a few percent of the gas reservoir. On the contrary, higher-mass seeds ($M_{\bullet}>M_{crit}$) would go through a feeding-dominated growth, with outflows playing a negligible role: the black hole grows in mass very rapidly, possibly even consuming most of the host halo mass, reaching the SMBH stage early in time. For the LDP case, and for slim disk accretion in either density profile $M_{crit}$ is very low ($\sim 10-100 \, \mathrm{\Msun}$), and therefore of relevance only if black hole seeds are stellar-mass or so. For the HDP case, and standard accretion, $M_{crit} \sim 3 \times 10^6 \, \mathrm{\Msun}$ is relevant for most seed masses proposed in the literature.

\begin{figure}
\vspace{-1\baselineskip}
\hspace{-0.5cm}
\begin{center}
\includegraphics[angle=0,width=0.50\textwidth]{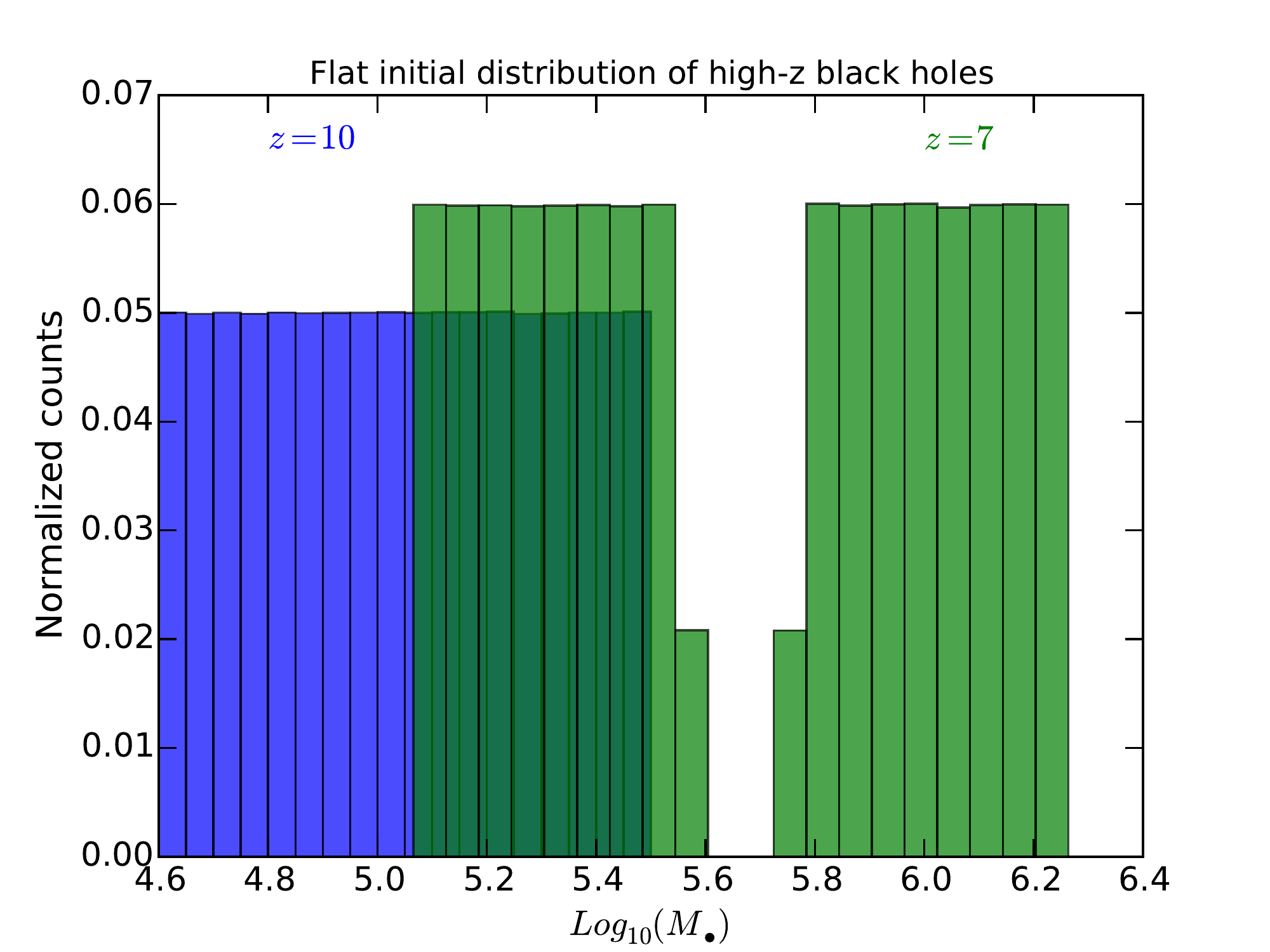}
\vspace{-0.1cm}
\includegraphics[angle=0,width=0.50\textwidth]{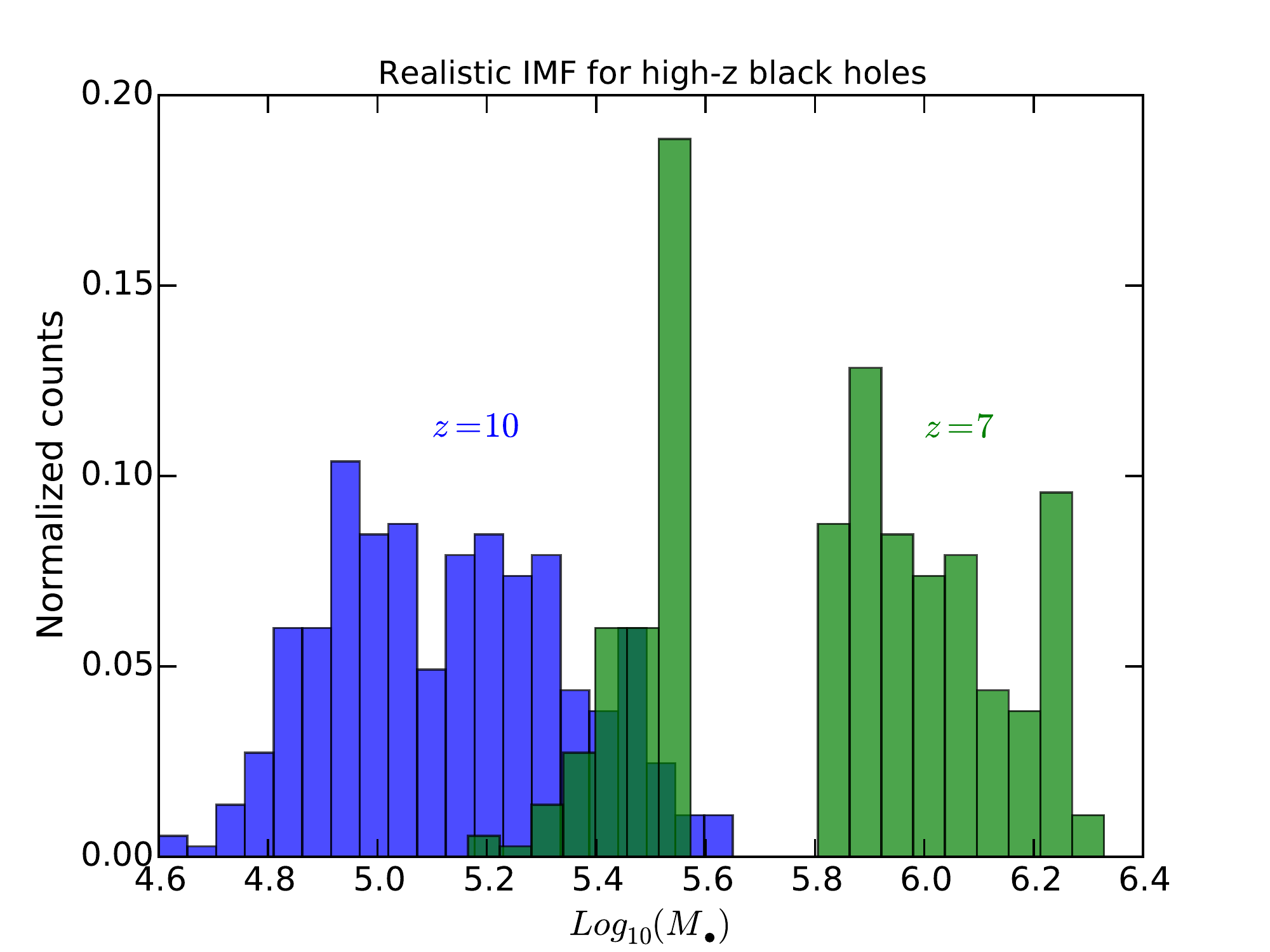}
\caption{Proof-of-concept bimodal evolution, between $z=10$ and $z=7$, of two initial mass functions for black hole seeds: a flat one (top) and a more realistic one (bottom). The evolution is computed along the theoretical lines described in this paper, with values for ${\cal D}(M_{\bullet})$ and $f_{Edd}(M_{\bullet})$ interpolated from the HDP simulations in the standard accretion scenario.}
\label{fig:IMF_evolution}
\end{center}
\end{figure}

As a proof-of-concept of this bimodal development, Fig. \ref{fig:IMF_evolution} shows the cosmological evolution, between $z=10$ and $z=7$ (the epoch when the first SMBHs are observed), of two initial mass functions for high-redshift black hole seeds: a simple flat distribution in the mass range $Log_{10}(M_{\bullet}[\mathrm{\Msun}])=4.5-5.5$ (top panel) and the initial mass function modeled (at $z=10$) in \cite{Ferrara_2014} (bottom panel). This evolution, far from being a precise prediction of the actual black hole growth, is a proof-of-concept based on the theoretical framework described in the present work.
The basic equation for the mass growth is the following one:
\begin{equation}
M_{\bullet}(t) = M_{\bullet}(t=0) \exp\left[ {\cal D}f_{Edd} \frac{t}{0.045 \, \mathrm{Gyr}} \right] \, ,
\end{equation}
where the values for ${\cal D}(M_{\bullet})$ and $f_{Edd}(M_{\bullet})$ are interpolated for each $M_{\bullet}$ from the solid lines in Fig. \ref{fig:comparison_standard}, in the HDP case. 
The bimodal evolution is evident and the mass gap at $Log_{10}(M_{\bullet}) \sim 5.7$ is expected to rapidly spread during the cosmic time. Importantly, this effect does not depend on the shape of the initial mass function.

In the HDP case with a radiatively inefficient accretion the bimodal evolution is expected to occur as well since, while ${\cal D} \sim 1$ for all masses, the value of $f_{Edd}$ does show a mass dependence (see the upper panel of Fig \ref{fig:comparison_SD}).

In halos with a LDP density profile, in any accretion scenario, the bimodal evolution is not expected to occur, since $f_{Edd}$ and ${\cal D}$ are nearly independent of the seed mass (see bottom panels of Figs. \ref{fig:comparison_standard} and \ref{fig:comparison_SD}). Nonetheless, we expect this evolutionary effect to play a remarkable role in the growth process, since we believe that dark matter halos with a HDP density profile harbored at their center the black hole seeds with smaller masses ($M_{\bullet} \lsim 10^{3-4} \, \mathrm{\Msun}$), the ones which are affected the most by feedback-limited growth.

\section{Discussion and Conclusions}
\label{sec:disc_concl}
The aim of this work is to provide a theoretical framework, supported by numerical simulations, to describe the growth of high-redshift ($z \sim 10$) black hole seeds.
The growth can be either feeding-dominated or feedback-limited. It is feeding-dominated if the radiative back-reaction of the black hole is negligible: the rapidity of the process is mainly determined by the gas accretion rate that the host halo can provide.
In a feeding-dominated accretion flow the black hole: (i) increases its mass during most of its time evolution, and (ii) exerts a relatively small mechanical and/or radiative feedback on its surrounding gas reservoir.
For these reasons, a feeding-dominated accretion is the desirable way to grow the black hole efficiently.

Within this theoretical framework, we investigated, with the aid of 1D radiation-hydrodynamic simulations, the growth of black hole seeds with a set of astrophysically-motivated initial conditions to explore the consequences of our model: seed masses in the range $10^{3-6} \, \mathrm{\Msun}$, embedded in a dark matter halo of total mass (dark matter and baryons) $M_h  = 6.7 \times 10^8 \, \mathrm{M}_{\odot}$ with two different density profiles and with different prescriptions for the accretion efficiency, namely a radiatively efficient mode ($\epsilon = 0.1$) and a slim disk mode ($\epsilon \lsim 0.04$).

Three points in particular are worthy of being retained from this work.
\begin{itemize}
\vspace{-2.0pt}
\item We confirmed that radiatively inefficient accretion modes (for instance the slim disk model) may ensure a continuous growth with rates largely exceeding the Eddington limit (reaching $\sim 300\dot{M}_{Edd}$ in our simulations). Radiatively inefficient accretion flows allow for feeding-dominated growths of the central black hole, while standard accretion scenarios may be feedback-limited with high values of the host halo gas density. The feedback time scale for these radiatively inefficient modes is $\gsim 60$ times longer than in the standard accretion scenario: the system reacts to a modification of the accretion rate in a much slower way because at a given accretion rate the production of radiation is reduced, decreasing the feedback effectiveness. In addition, we numerically proved the feasibility of accretion flows with \textit{sub-Eddington luminosities} and \textit{super-Eddington rates}.

\item We theoretically derived the existence of a time-evolving transition radius, $r_T$, which discriminates between feeding-dominated and feedback-limited growths. The transition radius, in addition, determines the spatial scale at which outflows take place, and provides a mass scale, $M_{crit}$, above which the black hole growth is always feeding-dominated.
The critical black hole mass is $10-10^6 \, \mathrm{\Msun}$, depending on the accretion scenario and on the host halo properties. Consequently, we foresee the possibility of a bimodal evolution of the population of black hole seeds:  low-mass seeds grow much less efficiently than high-mass ones. 

\item Our model may be employed in modeling the growth of high-redshift black holes in large cosmological simulations, which cannot resolve the typical spatial scales of accretion.
\end{itemize}
\vspace{-2.0pt}

To conclude, our model's aim is to study high-redshift accretion flows leading to the growth of the first black holes. In particular, we provided the theoretical framework needed to understand why radiatively inefficient accretion models are likely to be a crucial ingredient in explaining the presence of SMBHs of mass $\sim 10^{9-10} \, \mathrm{\Msun}$ less than $1 \, \mathrm{Gyr}$ after the Big Bang \citep{Mortlock_2011, Wu_2015}. In order to build such extremely massive objects already at $z \sim 6.5$, an Eddington-capped accretion would require, starting from a stellar mass ($M_{\bullet} \sim 50 \, \mathrm{\Msun}$) seed, a constant Eddington flow. However, a low-mass seed would be hindered in its initial growth by its own feedback, making continuous accretion at Eddington levels unlikely \citep[see also][]{JBromm,Alvarez2009,Milosavljevic_2009,Ricotti_2012_DC}. 

\section{Acknowledgements}
We thank Bin Yue for useful comments and suggestions. MV acknowledges support from a Marie Curie FP7-Reintegration-Grant  (PCIG10-GA-2011-303609).


\bibliographystyle{mnras}
\bibliography{ms}

\begin{thebibliography}{}
\makeatletter
\relax
\def\mn@urlcharsother{\let\do\@makeother \do\$\do\&\do\#\do\^\do\_\do\%\do\~}
\def\mn@doi{\begingroup\mn@urlcharsother \@ifnextchar [ {\mn@doi@}
  {\mn@doi@[]}}
\def\mn@doi@[#1]#2{\def\@tempa{#1}\ifx\@tempa\@empty \href
  {http://dx.doi.org/#2} {doi:#2}\else \href {http://dx.doi.org/#2} {#1}\fi
  \endgroup}
\def\mn@eprint#1#2{\mn@eprint@#1:#2::\@nil}
\def\mn@eprint@arXiv#1{\href {http://arxiv.org/abs/#1} {{\tt arXiv:#1}}}
\def\mn@eprint@dblp#1{\href {http://dblp.uni-trier.de/rec/bibtex/#1.xml}
  {dblp:#1}}
\def\mn@eprint@#1:#2:#3:#4\@nil{\def\@tempa {#1}\def\@tempb {#2}\def\@tempc
  {#3}\ifx \@tempc \@empty \let \@tempc \@tempb \let \@tempb \@tempa \fi \ifx
  \@tempb \@empty \def\@tempb {arXiv}\fi \@ifundefined
  {mn@eprint@\@tempb}{\@tempb:\@tempc}{\expandafter \expandafter \csname
  mn@eprint@\@tempb\endcsname \expandafter{\@tempc}}}

\bibitem[\protect\citeauthoryear{{Abramowicz} \& {Fragile}}{{Abramowicz} \&
  {Fragile}}{2013}]{2013LRR....16....1A}
{Abramowicz} M.~A.,  {Fragile} P.~C.,  2013, \mn@doi [Living Reviews in
  Relativity] {10.12942/lrr-2013-1}, \href
  {http://adsabs.harvard.edu/abs/2013LRR....16....1A} {16, 1}

\bibitem[\protect\citeauthoryear{{Abramowicz}, {Czerny}, {Lasota}  \&
  {Szuszkiewicz}}{{Abramowicz} et~al.}{1988}]{Abramowicz_1988}
{Abramowicz} M.~A.,  {Czerny} B.,  {Lasota} J.~P.,   {Szuszkiewicz} E.,  1988,
  \mn@doi [\apj] {10.1086/166683}, \href
  {http://adsabs.harvard.edu/abs/1988ApJ...332..646A} {332, 646}

\bibitem[\protect\citeauthoryear{{Alexander} \& {Natarajan}}{{Alexander} \&
  {Natarajan}}{2014}]{Alexander_2014}
{Alexander} T.,  {Natarajan} P.,  2014, \mn@doi [Science]
  {10.1126/science.1251053}, \href
  {http://adsabs.harvard.edu/abs/2014Sci...345.1330A} {345, 1330}

\bibitem[\protect\citeauthoryear{{Alvarez}, {Wise}  \& {Abel}}{{Alvarez}
  et~al.}{2009}]{Alvarez2009}
{Alvarez} M.~A.,  {Wise} J.~H.,   {Abel} T.,  2009, \mn@doi [\apjl]
  {10.1088/0004-637X/701/2/L133}, \href
  {http://esoads.eso.org/abs/2009ApJ...701L.133A} {701, L133}

\bibitem[\protect\citeauthoryear{{Begelman}}{{Begelman}}{1978}]{Begelman_1978}
{Begelman} M.~C.,  1978, \mnras, \href
  {http://adsabs.harvard.edu/abs/1978MNRAS.184...53B} {184, 53}

\bibitem[\protect\citeauthoryear{{Begelman}}{{Begelman}}{2012}]{2012MNRAS.420.2912B}
{Begelman} M.~C.,  2012, \mn@doi [\mnras] {10.1111/j.1365-2966.2011.20071.x},
  \href {http://adsabs.harvard.edu/abs/2012MNRAS.420.2912B} {420, 2912}

\bibitem[\protect\citeauthoryear{{Begelman}, {Volonteri}  \& {Rees}}{{Begelman}
  et~al.}{2006}]{Begelman_2006}
{Begelman} M.~C.,  {Volonteri} M.,   {Rees} M.~J.,  2006, \mn@doi [\mnras]
  {10.1111/j.1365-2966.2006.10467.x}, \href
  {http://adsabs.harvard.edu/abs/2006MNRAS.370..289B} {370, 289}

\bibitem[\protect\citeauthoryear{{Begelman}, {Rossi}  \& {Armitage}}{{Begelman}
  et~al.}{2008}]{Begelman_2008}
{Begelman} M.~C.,  {Rossi} E.~M.,   {Armitage} P.~J.,  2008, \mn@doi [\mnras]
  {10.1111/j.1365-2966.2008.13344.x}, \href
  {http://adsabs.harvard.edu/abs/2008MNRAS.387.1649B} {387, 1649}

\bibitem[\protect\citeauthoryear{{Blandford} \& {Begelman}}{{Blandford} \&
  {Begelman}}{1999}]{1999MNRAS.303L...1B}
{Blandford} R.~D.,  {Begelman} M.~C.,  1999, \mn@doi [\mnras]
  {10.1046/j.1365-8711.1999.02358.x}, \href
  {http://adsabs.harvard.edu/abs/1999MNRAS.303L...1B} {303, L1}

\bibitem[\protect\citeauthoryear{{Bondi}}{{Bondi}}{1952}]{Bondi_1952}
{Bondi} H.,  1952, \mnras, \href
  {http://adsabs.harvard.edu/abs/1952MNRAS.112..195B} {112, 195}

\bibitem[\protect\citeauthoryear{{Bromm} \& {Loeb}}{{Bromm} \&
  {Loeb}}{2003}]{Bromm_Loeb_2003}
{Bromm} V.,  {Loeb} A.,  2003, \mn@doi [\apj] {10.1086/377529}, \href
  {http://adsabs.harvard.edu/abs/2003ApJ...596...34B} {596, 34}

\bibitem[\protect\citeauthoryear{{Choi}, {Shlosman}  \& {Begelman}}{{Choi}
  et~al.}{2013}]{Choi_2013}
{Choi} J.-H.,  {Shlosman} I.,   {Begelman} M.~C.,  2013, \mn@doi [\apj]
  {10.1088/0004-637X/774/2/149}, \href
  {http://adsabs.harvard.edu/abs/2013ApJ...774..149C} {774, 149}

\bibitem[\protect\citeauthoryear{{Choi}, {Shlosman}  \& {Begelman}}{{Choi}
  et~al.}{2015}]{Choi_2015}
{Choi} J.-H.,  {Shlosman} I.,   {Begelman} M.~C.,  2015, \mn@doi [\mnras]
  {10.1093/mnras/stv694}, \href
  {http://adsabs.harvard.edu/abs/2015MNRAS.450.4411C} {450, 4411}

\bibitem[\protect\citeauthoryear{{Costa}, {Sijacki}, {Trenti}  \&
  {Haehnelt}}{{Costa} et~al.}{2014}]{Costa_2014}
{Costa} T.,  {Sijacki} D.,  {Trenti} M.,   {Haehnelt} M.~G.,  2014, \mn@doi
  [\mnras] {10.1093/mnras/stu101}, \href
  {http://adsabs.harvard.edu/abs/2014MNRAS.439.2146C} {439, 2146}

\bibitem[\protect\citeauthoryear{{Coughlin} \& {Begelman}}{{Coughlin} \&
  {Begelman}}{2014}]{Coughlin_2014}
{Coughlin} E.~R.,  {Begelman} M.~C.,  2014, \mn@doi [\apj]
  {10.1088/0004-637X/781/2/82}, \href
  {http://adsabs.harvard.edu/abs/2014ApJ...781...82C} {781, 82}

\bibitem[\protect\citeauthoryear{{Davies}, {Miller}  \& {Bellovary}}{{Davies}
  et~al.}{2011}]{Davies_2011}
{Davies} M.~B.,  {Miller} M.~C.,   {Bellovary} J.~M.,  2011, \mn@doi [\apjl]
  {10.1088/2041-8205/740/2/L42}, \href
  {http://adsabs.harvard.edu/abs/2011ApJ...740L..42D} {740, L42}

\bibitem[\protect\citeauthoryear{{Devecchi} \& {Volonteri}}{{Devecchi} \&
  {Volonteri}}{2009}]{Devecchi_2009}
{Devecchi} B.,  {Volonteri} M.,  2009, \mn@doi [\apj]
  {10.1088/0004-637X/694/1/302}, \href
  {http://adsabs.harvard.edu/abs/2009ApJ...694..302D} {694, 302}

\bibitem[\protect\citeauthoryear{{Di Matteo}, {Colberg}, {Springel},
  {Hernquist}  \& {Sijacki}}{{Di Matteo} et~al.}{2008}]{DiMatteo_2008}
{Di Matteo} T.,  {Colberg} J.,  {Springel} V.,  {Hernquist} L.,   {Sijacki} D.,
   2008, \mn@doi [\apj] {10.1086/524921}, \href
  {http://adsabs.harvard.edu/abs/2008ApJ...676...33D} {676, 33}

\bibitem[\protect\citeauthoryear{{Dubois}, {Pichon}, {Devriendt}, {Silk},
  {Haehnelt}, {Kimm}  \& {Slyz}}{{Dubois} et~al.}{2013}]{2013MNRAS.428.2885D}
{Dubois} Y.,  {Pichon} C.,  {Devriendt} J.,  {Silk} J.,  {Haehnelt} M.,  {Kimm}
  T.,   {Slyz} A.,  2013, \mn@doi [\mnras] {10.1093/mnras/sts224}, \href
  {http://adsabs.harvard.edu/abs/2013MNRAS.428.2885D} {428, 2885}

\bibitem[\protect\citeauthoryear{{Dubois}, {Volonteri}, {Silk}, {Devriendt},
  {Slyz}  \& {Teyssier}}{{Dubois} et~al.}{2015}]{2015arXiv150400018D}
{Dubois} Y.,  {Volonteri} M.,  {Silk} J.,  {Devriendt} J.,  {Slyz} A.,
  {Teyssier} R.,  2015, preprint, \href
  {http://adsabs.harvard.edu/abs/2015arXiv150400018D} {} (\mn@eprint {arXiv}
  {1504.00018})

\bibitem[\protect\citeauthoryear{{Fan} et~al.}{{Fan} et~al.}{2006}]{Fan_2006}
{Fan} X.,  et~al., 2006, \mn@doi [\aj] {10.1086/500296}, \href
  {http://adsabs.harvard.edu/abs/2006AJ....131.1203F} {131, 1203}

\bibitem[\protect\citeauthoryear{{Ferrara}, {Salvadori}, {Yue}  \&
  {Schleicher}}{{Ferrara} et~al.}{2014}]{Ferrara_2014}
{Ferrara} A.,  {Salvadori} S.,  {Yue} B.,   {Schleicher} D.,  2014, \mn@doi
  [\mnras] {10.1093/mnras/stu1280}, \href
  {http://adsabs.harvard.edu/abs/2014MNRAS.443.2410F} {443, 2410}

\bibitem[\protect\citeauthoryear{{Haiman}}{{Haiman}}{2013}]{Haiman_2013}
{Haiman} Z.,  2013, in {Wiklind} T.,  {Mobasher} B.,   {Bromm} V.,  eds,
  Astrophysics and Space Science Library Vol. 396, Astrophysics and Space
  Science Library. p.~293 (\mn@eprint {arXiv} {1203.6075}),
  \mn@doi{10.1007/978-3-642-32362-1_6}

\bibitem[\protect\citeauthoryear{{Jeon}, {Pawlik}, {Greif}, {Glover}, {Bromm},
  {Milosavljevi{\'c}}  \& {Klessen}}{{Jeon} et~al.}{2012}]{Jeon_2012}
{Jeon} M.,  {Pawlik} A.~H.,  {Greif} T.~H.,  {Glover} S.~C.~O.,  {Bromm} V.,
  {Milosavljevi{\'c}} M.,   {Klessen} R.~S.,  2012, \mn@doi [\apj]
  {10.1088/0004-637X/754/1/34}, \href
  {http://adsabs.harvard.edu/abs/2012ApJ...754...34J} {754, 34}

\bibitem[\protect\citeauthoryear{{Jiang}, {Stone}  \& {Davis}}{{Jiang}
  et~al.}{2014}]{Jiang_2014}
{Jiang} Y.-F.,  {Stone} J.~M.,   {Davis} S.~W.,  2014, \mn@doi [\apj]
  {10.1088/0004-637X/796/2/106}, \href
  {http://adsabs.harvard.edu/abs/2014ApJ...796..106J} {796, 106}

\bibitem[\protect\citeauthoryear{{Johnson} \& {Bromm}}{{Johnson} \&
  {Bromm}}{2007}]{JBromm}
{Johnson} J.~L.,  {Bromm} V.,  2007, \mn@doi [MNRAS]
  {10.1111/j.1365-2966.2006.11275.x}, \href
  {http://adsabs.harvard.edu/abs/2007MNRAS.374.1557J} {374, 1557}

\bibitem[\protect\citeauthoryear{{Johnson}, {Whalen}, {Fryer}  \&
  {Li}}{{Johnson} et~al.}{2012}]{Johnson_2012}
{Johnson} J.~L.,  {Whalen} D.~J.,  {Fryer} C.~L.,   {Li} H.,  2012, \mn@doi
  [\apj] {10.1088/0004-637X/750/1/66}, \href
  {http://adsabs.harvard.edu/abs/2012ApJ...750...66J} {750, 66}

\bibitem[\protect\citeauthoryear{{King}}{{King}}{2003}]{King_2003}
{King} A.,  2003, \mn@doi [\apjl] {10.1086/379143}, \href
  {http://adsabs.harvard.edu/abs/2003ApJ...596L..27K} {596, L27}

\bibitem[\protect\citeauthoryear{{King}}{{King}}{2010}]{King_2010}
{King} A.~R.,  2010, \mn@doi [\mnras] {10.1111/j.1365-2966.2009.16013.x}, \href
  {http://adsabs.harvard.edu/abs/2010MNRAS.402.1516K} {402, 1516}

\bibitem[\protect\citeauthoryear{{Lasota}}{{Lasota}}{2015}]{2015arXiv150502172L}
{Lasota} J.-P.,  2015, preprint, \href
  {http://adsabs.harvard.edu/abs/2015arXiv150502172L} {} (\mn@eprint {arXiv}
  {1505.02172})

\bibitem[\protect\citeauthoryear{{Latif}, {Schleicher}, {Schmidt}  \&
  {Niemeyer}}{{Latif} et~al.}{2013a}]{Latif_2013}
{Latif} M.~A.,  {Schleicher} D.~R.~G.,  {Schmidt} W.,   {Niemeyer} J.,  2013a,
  \mn@doi [\mnras] {10.1093/mnras/stt834}, \href
  {http://adsabs.harvard.edu/abs/2013MNRAS.433.1607L} {433, 1607}

\bibitem[\protect\citeauthoryear{{Latif}, {Schleicher}, {Schmidt}  \&
  {Niemeyer}}{{Latif} et~al.}{2013b}]{Latif_2013c}
{Latif} M.~A.,  {Schleicher} D.~R.~G.,  {Schmidt} W.,   {Niemeyer} J.~C.,
  2013b, \mn@doi [\mnras] {10.1093/mnras/stt1786}, \href
  {http://adsabs.harvard.edu/abs/2013MNRAS.436.2989L} {436, 2989}

\bibitem[\protect\citeauthoryear{{Latif}, {Niemeyer}  \& {Schleicher}}{{Latif}
  et~al.}{2014}]{Latif_2014b}
{Latif} M.~A.,  {Niemeyer} J.~C.,   {Schleicher} D.~R.~G.,  2014, \mn@doi
  [\mnras] {10.1093/mnras/stu489}, \href
  {http://adsabs.harvard.edu/abs/2014MNRAS.440.2969L} {440, 2969}

\bibitem[\protect\citeauthoryear{{Lodato} \& {Natarajan}}{{Lodato} \&
  {Natarajan}}{2006}]{Lodato_Natarajan_2006}
{Lodato} G.,  {Natarajan} P.,  2006, \mn@doi [\mnras]
  {10.1111/j.1365-2966.2006.10801.x}, \href
  {http://adsabs.harvard.edu/abs/2006MNRAS.371.1813L} {371, 1813}

\bibitem[\protect\citeauthoryear{{Lodato} \& {Pringle}}{{Lodato} \&
  {Pringle}}{2007}]{Lodato_2007}
{Lodato} G.,  {Pringle} J.~E.,  2007, \mn@doi [\mnras]
  {10.1111/j.1365-2966.2007.12332.x}, \href
  {http://adsabs.harvard.edu/abs/2007MNRAS.381.1287L} {381, 1287}

\bibitem[\protect\citeauthoryear{{Madau} \& {Rees}}{{Madau} \&
  {Rees}}{2001}]{Madau_Rees_2001}
{Madau} P.,  {Rees} M.~J.,  2001, \mn@doi [\apjl] {10.1086/319848}, \href
  {http://adsabs.harvard.edu/abs/2001ApJ...551L..27M} {551, L27}

\bibitem[\protect\citeauthoryear{{Madau}, {Haardt}  \& {Dotti}}{{Madau}
  et~al.}{2014}]{Madau_2014}
{Madau} P.,  {Haardt} F.,   {Dotti} M.,  2014, \mn@doi [\apjl]
  {10.1088/2041-8205/784/2/L38}, \href
  {http://adsabs.harvard.edu/abs/2014ApJ...784L..38M} {784, L38}

\bibitem[\protect\citeauthoryear{{McKinney}, {Tchekhovskoy}, {Sadowski}  \&
  {Narayan}}{{McKinney} et~al.}{2014}]{McKinney_2014}
{McKinney} J.~C.,  {Tchekhovskoy} A.,  {Sadowski} A.,   {Narayan} R.,  2014,
  \mn@doi [\mnras] {10.1093/mnras/stu762}, \href
  {http://adsabs.harvard.edu/abs/2014MNRAS.441.3177M} {441, 3177}

\bibitem[\protect\citeauthoryear{{Milosavljevi{\'c}}, {Bromm}, {Couch}  \&
  {Oh}}{{Milosavljevi{\'c}} et~al.}{2009}]{Milosavljevic_2009}
{Milosavljevi{\'c}} M.,  {Bromm} V.,  {Couch} S.~M.,   {Oh} S.~P.,  2009,
  \mn@doi [\apj] {10.1088/0004-637X/698/1/766}, \href
  {http://adsabs.harvard.edu/abs/2009ApJ...698..766M} {698, 766}

\bibitem[\protect\citeauthoryear{{Mineshige}, {Kawaguchi}, {Takeuchi}  \&
  {Hayashida}}{{Mineshige} et~al.}{2000}]{Mineshige_2000}
{Mineshige} S.,  {Kawaguchi} T.,  {Takeuchi} M.,   {Hayashida} K.,  2000,
  \mn@doi [\pasj] {10.1093/pasj/52.3.499}, \href
  {http://adsabs.harvard.edu/abs/2000PASJ...52..499M} {52, 499}

\bibitem[\protect\citeauthoryear{{Mortlock} et~al.,}{{Mortlock}
  et~al.}{2011}]{Mortlock_2011}
{Mortlock} D.~J.,  et~al., 2011, \mn@doi [\nat] {10.1038/nature10159}, \href
  {http://adsabs.harvard.edu/abs/2011Natur.474..616M} {474, 616}

\bibitem[\protect\citeauthoryear{{Novak}, {Ostriker}  \& {Ciotti}}{{Novak}
  et~al.}{2011}]{Novak_2011}
{Novak} G.~S.,  {Ostriker} J.~P.,   {Ciotti} L.,  2011, \mn@doi [\apj]
  {10.1088/0004-637X/737/1/26}, \href
  {http://adsabs.harvard.edu/abs/2011ApJ...737...26N} {737, 26}

\bibitem[\protect\citeauthoryear{{Novak}, {Ostriker}  \& {Ciotti}}{{Novak}
  et~al.}{2012}]{Novak_2012}
{Novak} G.~S.,  {Ostriker} J.~P.,   {Ciotti} L.,  2012, \mn@doi [\mnras]
  {10.1111/j.1365-2966.2012.21844.x}, \href
  {http://adsabs.harvard.edu/abs/2012MNRAS.427.2734N} {427, 2734}

\bibitem[\protect\citeauthoryear{{Ohsuga}, {Mineshige}, {Mori}  \&
  {Umemura}}{{Ohsuga} et~al.}{2002}]{Ohsuga_2002}
{Ohsuga} K.,  {Mineshige} S.,  {Mori} M.,   {Umemura} M.,  2002, \mn@doi [\apj]
  {10.1086/340798}, \href {http://adsabs.harvard.edu/abs/2002ApJ...574..315O}
  {574, 315}

\bibitem[\protect\citeauthoryear{{Pacucci} \& {Ferrara}}{{Pacucci} \&
  {Ferrara}}{2015}]{Pacucci_2015}
{Pacucci} F.,  {Ferrara} A.,  2015, \mn@doi [\mnras] {10.1093/mnras/stv018},
  \href {http://adsabs.harvard.edu/abs/2015MNRAS.448..104P} {448, 104}

\bibitem[\protect\citeauthoryear{{Pacucci}, {Ferrara}, {Volonteri}  \&
  {Dubus}}{{Pacucci} et~al.}{2015}]{Pacucci_F_V_2015}
{Pacucci} F.,  {Ferrara} A.,  {Volonteri} M.,   {Dubus} G.,  2015, preprint,
  \href {http://adsabs.harvard.edu/abs/2015arXiv150605299P} {} (\mn@eprint
  {arXiv} {1506.05299})

\bibitem[\protect\citeauthoryear{{Paczynski} \& {Abramowicz}}{{Paczynski} \&
  {Abramowicz}}{1982}]{Paczynski_1982}
{Paczynski} B.,  {Abramowicz} M.~A.,  1982, \mn@doi [\apj] {10.1086/159689},
  \href {http://adsabs.harvard.edu/abs/1982ApJ...253..897P} {253, 897}

\bibitem[\protect\citeauthoryear{{Park} \& {Ricotti}}{{Park} \&
  {Ricotti}}{2012}]{Ricotti_2012_DC}
{Park} K.,  {Ricotti} M.,  2012, \mn@doi [\apj] {10.1088/0004-637X/747/1/9},
  \href {http://adsabs.harvard.edu/abs/2012ApJ...747....9P} {747, 9}

\bibitem[\protect\citeauthoryear{{Petri}, {Ferrara}  \& {Salvaterra}}{{Petri}
  et~al.}{2012}]{Petri_2012}
{Petri} A.,  {Ferrara} A.,   {Salvaterra} R.,  2012, \mn@doi [\mnras]
  {10.1111/j.1365-2966.2012.20743.x}, \href
  {http://adsabs.harvard.edu/abs/2012MNRAS.422.1690P} {422, 1690}

\bibitem[\protect\citeauthoryear{{Planck Collaboration} et~al.,}{{Planck
  Collaboration} et~al.}{2015}]{Planck_2015}
{Planck Collaboration} et~al., 2015, ArXiv e-prints 1502.01589, \href
  {http://adsabs.harvard.edu/abs/2015arXiv150201589P} {}

\bibitem[\protect\citeauthoryear{{Sadowski}}{{Sadowski}}{2009}]{Sadowski_2009}
{Sadowski} A.,  2009, \mn@doi [\apjs] {10.1088/0067-0049/183/2/171}, \href
  {http://adsabs.harvard.edu/abs/2009ApJS..183..171S} {183, 171}

\bibitem[\protect\citeauthoryear{{Sadowski}}{{Sadowski}}{2011}]{Sadowski_2011}
{Sadowski} A.,  2011, ArXiv e-prints 1108.0396, \href
  {http://adsabs.harvard.edu/abs/2011arXiv1108.0396S} {}

\bibitem[\protect\citeauthoryear{{Sadowski}, {Narayan}, {McKinney}  \&
  {Tchekhovskoy}}{{Sadowski} et~al.}{2014}]{Sadowski_2014_b}
{Sadowski} A.,  {Narayan} R.,  {McKinney} J.~C.,   {Tchekhovskoy} A.,  2014,
  \mn@doi [\mnras] {10.1093/mnras/stt2479}, \href
  {http://adsabs.harvard.edu/abs/2014MNRAS.439..503S} {439, 503}

\bibitem[\protect\citeauthoryear{{Shakura} \& {Sunyaev}}{{Shakura} \&
  {Sunyaev}}{1973}]{Shakura_Sunyaev_1973}
{Shakura} N.~I.,  {Sunyaev} R.~A.,  1973, \aap, \href
  {http://adsabs.harvard.edu/abs/1973A%26A....24..337S} {24, 337}

\bibitem[\protect\citeauthoryear{{Shang}, {Bryan}  \& {Haiman}}{{Shang}
  et~al.}{2010}]{Shang_2010}
{Shang} C.,  {Bryan} G.~L.,   {Haiman} Z.,  2010, \mn@doi [\mnras]
  {10.1111/j.1365-2966.2009.15960.x}, \href
  {http://adsabs.harvard.edu/abs/2010MNRAS.402.1249S} {402, 1249}

\bibitem[\protect\citeauthoryear{{Silk} \& {Rees}}{{Silk} \&
  {Rees}}{1998}]{Silk_1998}
{Silk} J.,  {Rees} M.~J.,  1998, \aap, \href
  {http://adsabs.harvard.edu/abs/1998A%26A...331L...1S} {331, L1}

\bibitem[\protect\citeauthoryear{{Springel} et~al.,}{{Springel}
  et~al.}{2005}]{Springel_2005}
{Springel} V.,  et~al., 2005, \mn@doi [\nat] {10.1038/nature03597}, \href
  {http://adsabs.harvard.edu/abs/2005Natur.435..629S} {435, 629}

\bibitem[\protect\citeauthoryear{{Thorne}}{{Thorne}}{1974}]{Thorne_1974}
{Thorne} K.~S.,  1974, \mn@doi [\apj] {10.1086/152991}, \href
  {http://adsabs.harvard.edu/abs/1974ApJ...191..507T} {191, 507}

\bibitem[\protect\citeauthoryear{{Volonteri}}{{Volonteri}}{2010}]{Volonteri_2010}
{Volonteri} M.,  2010, \mn@doi [{Astronomy and Astrophysics Review}]
  {10.1007/s00159-010-0029-x}, \href
  {http://adsabs.harvard.edu/abs/2010A%26ARv..18..279V} {18, 279}

\bibitem[\protect\citeauthoryear{{Volonteri} \& {Rees}}{{Volonteri} \&
  {Rees}}{2005}]{Volonteri_2005}
{Volonteri} M.,  {Rees} M.~J.,  2005, \mn@doi [\apj] {10.1086/466521}, \href
  {http://adsabs.harvard.edu/abs/2005ApJ...633..624V} {633, 624}

\bibitem[\protect\citeauthoryear{{Volonteri}, {Silk}  \& {Dubus}}{{Volonteri}
  et~al.}{2014}]{Volonteri_2014}
{Volonteri} M.,  {Silk} J.,   {Dubus} G.,  2014, ArXiv e-prints 1401.3513,
  \href {http://adsabs.harvard.edu/abs/2014arXiv1401.3513V} {}

\bibitem[\protect\citeauthoryear{Wu et~al.,}{Wu et~al.}{2015}]{Wu_2015}
Wu X.-B.,  et~al., 2015, Nature, 518, 512

\makeatother
\end{thebibliography}

\label{lastpage}
\end{document}